\documentclass[useAMS,useusegraphicx]{mn2e}
\usepackage{graphicx}
\usepackage{times}

\newcommand\arcdeg{\mbox{$^\circ$}}%
\newcommand\lesssim{\mbox{$\let\la$}}%
\newcommand\gtrsim{\mbox{$\let\ga$}}%
\newcommand\aj{AJ}%
\newcommand\apj{ApJ}%
\newcommand\apjl{ApJ}%
\newcommand\apjs{ApJS}%
\newcommand\aap{A\&A}%
\newcommand\mnras{MNRAS}%

\title[The 2dF-SDSS LRG and QSO Survey: The Quasar Luminosity Function]{The 2dF-SDSS LRG and QSO Survey: The $z<2.1$ Quasar Luminosity Function from 5645 Quasars to $g=21.85$}

\author[G.~T. Richards et al.]
{Gordon T. Richards,$^1$
Scott M. Croom,$^2$
Scott F. Anderson,$^3$
Joss Bland-Hawthorn,$^2$
\newauthor
Brian J. Boyle,$^4$
Roberto De Propris,$^5$
Michael J. Drinkwater,$^6$
Xiaohui Fan,$^7$
\newauthor
James E. Gunn,$^1$
\v{Z}eljko Ivezi\'{c},$^{1,3}$
Sebastian Jester,$^8$
Jon Loveday,$^9$
Avery Meiksin,$^{10}$
\newauthor
Lance Miller,$^{11}$
Adam Myers,$^{12}$
Robert C. Nichol,$^{13}$
Phil J. Outram,$^{14}$
\newauthor
Kevin A. Pimbblet,$^6$
Isaac G. Roseboom,$^6$
Nic Ross,$^{14}$
Donald P. Schneider,$^{15}$
\newauthor
Tom Shanks,$^{14}$
Robert G. Sharp,$^2$
Chris Stoughton,$^8$
Michael A. Strauss,$^1$
\newauthor
Alexander S. Szalay,$^{16}$
Daniel E. Vanden Berk,$^{15}$
and Donald G. York$^{17,18}$
\\
$^1$Princeton University Observatory, Peyton Hall, Princeton, NJ 08544, USA\\
$^2$Anglo-Australian Observatory, PO Box 296, Epping, NSW 1710, Australia\\
$^3$Department of Astronomy, University of Washington, Box 351580, Seattle, WA 98195, USA\\
$^4$Australia Telescope National Facility, PO Box 76, Epping NSW 1710, Australia\\
$^5$H.H. Wills Physics Laboratory, University of Bristol, Tyndall Avenue, Bristol BS8 1TL\\
$^6$Department of Physics, University of Queensland, Brisbane, QLD 4072, Australia\\
$^7$Steward Observatory, University of Arizona, 933 North Cherry Avenue, Tucson, AZ 85721, USA\\
$^8$Fermi National Accelerator Laboratory, PO Box 500, Batavia, IL 60510, USA\\
$^9$Astronomy Centre, University of Sussex, Falmer, Brighton BN1 9QJ\\
$^{10}$Institute for Astronomy, Royal Observatory, University of Edinburgh, Blackford Hill, Edinburgh EH9 3HJ\\
$^{11}$Department of Physics, Oxford University, 1 Keble Road, Oxford OX1 3RH\\
$^{12}$Department of Astronomy, University of Illinois at Urbana-Champaign, 1002 West Green Street, Urbana, IL 61801-3080, USA\\
$^{13}$Institute of Cosmology and Gravitation, Mercantile House, Hampshire Terrace, University of Portsmouth, Portsmouth, PO1 2EG\\
$^{14}$Department of Physics, University of Durham, South Road, Durham DH1 3LE\\
$^{15}$Department of Astronomy and Astrophysics, The Pennsylvania State University, 525 Davey Laboratory, University Park, PA 16802, USA\\
$^{16}$Department of Physics and Astronomy, The Johns Hopkins University, 3400 North Charles Street, Baltimore, MD 21218-2686, USA\\
$^{17}$Department of Astronomy and Astrophysics, The University of Chicago, 5640 South Ellis Avenue, Chicago, IL 60637, USA\\
$^{18}$Enrico Fermi Institute, The University of Chicago, 5640 South Ellis Avenue, Chicago, IL 60637, USA\\
}

\begin{document}

\date{Accepted . Received ; in original form }

\pagerange{\pageref{firstpage}--\pageref{lastpage}} \pubyear{2005}

\maketitle

\label{firstpage}

\begin{abstract}
We have used the 2-degree Field (2dF) instrument on the
Anglo-Australian Telescope to obtain redshifts of a sample of $z<3$,
$18.0<g<21.85$ quasars selected from Sloan Digital Sky Survey (SDSS)
imaging.  These data are part of a larger joint programme between the
SDSS and 2dF communities to obtain spectra of faint quasars and
luminous red galaxies, namely the 2dF-SDSS LRG and QSO Survey (2SLAQ).
We describe the quasar selection algorithm and present the resulting
number counts and luminosity function of 5645 quasars in 105.7
deg$^2$.  The bright end number counts and luminosity function agree
well with determinations from the 2dF QSO Redshift Survey (2QZ) data
to $g\sim20.2$.  However, at the faint end the 2SLAQ number counts and
luminosity function are steeper (i.e. require more faint quasars) than
the final 2QZ results from Croom et al.\ (2004), but are consistent
with the preliminary 2QZ results from Boyle et al.\ (2000).  Using the
functional form adopted for the 2QZ analysis (a double-power law with
pure luminosity evolution characterized by a 2nd order polynomial in
redshift), we find a faint end slope of $\beta=-1.78\pm0.03$ if we
allow all of the parameters to vary and $\beta=-1.45\pm0.03$ if we
allow only the faint end slope and normalization to vary (holding all
other parameters equal to the final 2QZ values).  Over the magnitude
range covered by the 2SLAQ survey, our maximum likelihood fit to the
data yields 32 per cent more quasars than the final 2QZ
parameterization, but is not inconsistent with other $g>21$ deep
surveys for quasars.  The 2SLAQ data exhibit no well defined ``break''
in the number counts or luminosity function, but do clearly flatten
with increasing magnitude.  Finally, we find that the shape of the
quasar luminosity function derived from 2SLAQ is in good agreement
with that derived from type I quasars found in hard X-ray surveys.

\end{abstract}


\section{Introduction}

We have merged the high-quality digital imaging of the Sloan Digital
Sky Survey (SDSS; \nocite{yaa+00}{York} {et~al.} 2000) and the powerful spectroscopic
capabilities of the 2-degree Field (2dF) instrument \nocite{lct+02}({Lewis} {et~al.} 2002) to
conduct a deep wide-field spectroscopic survey for quasars and
luminous red galaxies (LRGs), i.e. the 2dF-SDSS LRG and QSO Survey
(hereafter referred to as ``2SLAQ'').  The combination of these
facilities allows us to probe substantially deeper than either the
SDSS or 2dF surveys can individually.  This paper describes the first
results of the quasar aspect of the survey; see \nocite{ccp+03}{Cannon et al.} (2003),
\nocite{pbs+04}{Padmanabhan et al.} (2004) and \nocite{can+05}{Cannon et al.} (2005) for a discussion of the LRG
component of the survey.  

The 2dF QSO Redshift survey (2QZ; \nocite{bsc+00,csb+04}{Boyle} {et~al.} 2000; {Croom} {et~al.} 2004) was
restricted to $b_J<20.85$.
We use the SDSS imaging data as the input for a
new survey, allowing us to probe to $g=21.85$ with typical photometric
errors at the flux limit of only 7 per cent -- considerably fainter than
either the $i=19.1$ flux limit of the SDSS Quasar Survey
\nocite{rfn+02,sfh+03}({Richards} {et~al.} 2002; {Schneider} {et~al.} 2003) or the $b_J=20.85$ flux limits of 2QZ.  By
allocating 200 fibres per 2dF plate to this new quasar survey (with an
additional 200 fibres going to LRGs), and extending the exposure time
to 4 hours (compared to $\sim1$ hour for SDSS and 2QZ), we hope to
obtain spectra of 10,000 quasars to $g=21.85$ in the next few years.
This paper reports on the first three semesters of data (with 5645
quasars) and presents the $z<2.1$ quasar luminosity function (QLF) to
fainter luminosities at each redshift than either the SDSS or 2QZ
surveys alone.

The first determination of the luminosity function of quasars was by
\nocite{sch68}{Schmidt} (1968).  Subsequent pioneering work was carried out by many
groups including \nocite{sg83}{Schmidt} \& {Green} (1983), \nocite{kk88}{Koo} \& {Kron} (1988), \nocite{hfc93}{Hewett}, {Foltz} \& {Chaffee} (1993) and
especially Boyle and collaborators \nocite{bsp88}(e.g. {Boyle}, {Shanks} \& {Peterson} 1988), with
extensions to high-$z$ ($z>3$) being provided by \nocite{who94}{Warren}, {Hewett} \& {Osmer} (1994),
\nocite{ssg95}{Schmidt}, {Schneider} \& {Gunn} (1995), \nocite{kdd+95}{Kennefick}, {Djorgovski} \& {de  Carvalho} (1995) and \nocite{fss+01}{Fan} {et~al.} (2001b).  The largest samples
analysed to date come from the 2dF QSO Redshift Survey
\nocite{bsc+00,csb+04}({Boyle} {et~al.} 2000; {Croom} {et~al.} 2004) with 23,338 quasars.

With the exception of variability-selected samples
\nocite{hv95}(e.g. {Hawkins} \& {Veron} 1995), early QLF determinations were generally
characterized by a strong, distinct ``break'' whose redshift evolution
has been the subject of much discussion.  However, more recent
determinations (e.g. COMBO-17; \nocite{wwb+03}{Wolf} {et~al.} 2003), while still
exhibiting distinct curvature in a log-log plot, show less of a break
at a specific luminosity.

Recently, optical surveys have been supplemented by X-ray surveys
(both soft and hard) that, when correcting for selection differences,
can largely reproduce the optical type I QLF
\nocite{uao+03,bcm+05}(e.g. {Ueda} {et~al.} 2003; {Barger et al.} 2005).  These X-ray QLFs have also been shown
to exhibit a break, but generally at luminosities much fainter than
found by optical surveys; this result suggests incompleteness at the
faint end of optical surveys.

As we shall see, our data are in good agreement with recent results
for faint quasars from both the optical \nocite{wwb+03}(e.g. {Wolf} {et~al.} 2003) and
X-ray \nocite{bcm+05}(e.g. {Barger et al.} 2005).  We probe nearly 1 magnitude deeper
than 2QZ, and find that the faint end slope of the QLF is steeper than
that of the most recent 2QZ determination and lacks a strong
characteristic break feature, but is still better characterized by a
double-power law than a single power-law.

Section~2 presents a description of the imaging data and the sample
selection.  In Section~3 we describe the observations and data
reduction.  Section~4 presents the completeness corrections leading to
the QLF presented in in \S~5.  Finally, \S~6 presents a discussion of
the ramifications of our work and summarizes our results.  Throughout
this paper we use a cosmology with $H_o=h_{70}70$\,km\,s$^{-1}$\,Mpc$^{-1}$,
$\Omega_m=0.3$, $\Omega_{\Lambda}=0.7$ \nocite{svp+03}(e.g. {Spergel} {et~al.} 2003).

\section{The imaging data and sample selection}
\label{sec:sample}

\subsection{The SDSS imaging data}

The photometric measurements used as the basis for our catalogue are
drawn from SDSS imaging data (DR1 reductions;
\nocite{slb+02,aaa+03}{Stoughton} {et~al.} 2002; {Abazajian} {et~al.} 2003), which will eventually cover roughly 10,000
deg$^{2}$ of sky in five photometric passbands ($ugriz$) using a
large-format charge-coupled device (CCD) camera \nocite{gcr+98}({Gunn} {et~al.} 1998).  The
photometric system and its characterization are discussed by
\nocite{fig+96}{Fukugita} {et~al.} (1996), \nocite{hfs+01}{Hogg} {et~al.} (2001), \nocite{stk+02}{Smith} {et~al.} (2002) and \nocite{slb+02}{Stoughton} {et~al.} (2002); the
spectroscopic tiling algorithm is described by \nocite{blm+03}{Blanton} {et~al.} (2003).  Except
where otherwise stated, all SDSS magnitudes discussed herein are
``asinh'' point-spread-function (PSF) magnitudes \nocite{lgs+99}({Lupton}, {Gunn} \& {Szalay} 1999) on an
AB magnitude system \nocite{og83}({Oke} \& {Gunn} 1983) that have been dereddened for
Galactic extinction according to the model of \nocite{sfd98}{Schlegel}, {Finkbeiner} \&  {Davis} (1998).  The
astrometric accuracy of the SDSS imaging data is better than 100 mas
per coordinate rms \nocite{pmh+03}({Pier} {et~al.} 2003).  The SDSS Quasar Survey
\nocite{rfn+02,sfh+03,sch+05}({Richards} {et~al.} 2002; {Schneider} {et~al.} 2003, 2005) extends to $i=19.1$ for $z<3$ and
$i=20.2$ for $z>3$, whereas our work herein explores the $z<3$ regime
to $g=21.85$ ($i\sim21.63$).

\subsection{Preliminary sample restrictions}

Our quasar candidate sample was drawn from 10 SDSS imaging runs (see
\S~2.4) after having first been vetted of objects that have cosmetic
defects (e.g. bad columns) that might cause the photometry to be
inaccurate.  Specifically, we rejected any objects that met the
``fatal'' or ``non-fatal'' error definitions as described by
\nocite{rfn+02}{Richards} {et~al.} (2002).

We next imposed limits on the $i$-band PSF magnitude and its estimated
$1\sigma$ error of $i<22.0$ and $\sigma_i<0.2$.  Further magnitude
cuts are done in the $g$-band (to facilitate comparisons with previous
2QZ results in $b_J$); the $i$-band cuts are primarily to reduce the
number of objects that we have to examine initially.  We also placed
restrictions on the errors in each of the other four bands,
specifically, $\sigma_u<0.4$, $\sigma_g<0.13$, $\sigma_r<0.13$ and
$\sigma_z<0.6$.  These restrictions are designed to ensure that the
errors on the magnitudes are reasonably small (and thus that the
resulting colours are accurate), but also are sufficiently relaxed
that, when coupled with the magnitude cut in $i$ and $g$, objects with
quasar-like continua are not rejected.  This tolerance is necessary
since, as we go fainter, restrictions on magnitude errors are
effectively cuts in magnitude and any two such restrictions are
effectively colour cuts.  Note that this selection of error
constraints effectively limits the redshift to less than 3, as the
Ly$\alpha$ forest suppresses the $u$ flux at higher redshifts.

\subsection{Colour cuts}

Based on spectroscopic identifications from SDSS and 2QZ of this
initial set of objects, we implement additional colour cuts that are
designed to efficiently select faint quasars while maintaining a high
degree of completeness to known UV-excess broad-line quasars.  An
analysis of the completeness of the selection algorithm is given as a
function of redshift and magnitude in \S~\ref{sec:photcomp}.

We first impose colour restrictions that are designed to reject hot
white dwarfs.  These cuts are made regardless of magnitude.
Specifically, we rejected objects that satisfy the condition:
$A\&\&((B\&\&C\&\&D)||E)$, where the letters refer to the cuts:
\begin{eqnarray}
\begin{array}{rrcccl}
{\rm A}) \;\, & -1.0 & < & u-g & < & 0.8 \\
{\rm B}) \;\, & -0.8 & < & g-r & < & 0.0 \\
{\rm C}) \;\, & -0.6 & < & r-i & < & -0.1 \\
{\rm D}) \;\, & -1.0 & < & i-z & < & -0.1 \\
{\rm E}) \;\, & -1.5 & < & g-i & < & -0.3. 
\end{array}
\end{eqnarray}
This is similar to the white dwarf cut applied by
\nocite{rfn+02}{Richards} {et~al.} (2002, Equation~2) except for the added cut with respect to
the $g-i$ colour.

As the targets become fainter and the magnitude errors increase, we
find that maximizing our completeness and efficiency is best served by
separate handling of bright and faint objects.  The bright sample is
restricted to $18.0<g<21.15$ and is designed to allow for overlap with
previous SDSS and 2dF spectroscopic observations.  The faint sample
has $21.15\le g<21.85$ and probes roughly one magnitude deeper than
2QZ.  These cuts are made in $g$ rather than $i$ (as the SDSS quasar
survey does) since we are concentrating on UV-excess quasars and would
like to facilitate comparison with the results from the $b_J$-based
2QZ.  The combination of the $g<21.85$ and $i<22.0$ cut will exclude
objects bluer than $\alpha_{\nu}=+0.3\; (f_{\nu}\propto\nu^{\alpha})$;
however, objects this blue are exceedingly rare ($>3\sigma$
deviations).

Further cuts are made as a function of colour and morphology in each
of the bright/faint samples.  In general, we would prefer not to make
a cut on morphology since we do not want to exclude low-$z$ quasars
and because our selection extends beyond the magnitude limits at which
the SDSS's star/galaxy separation breaks down.  However,
\nocite{sjd+02}{Scranton} {et~al.} (2002) have developed a Bayesian star-galaxy classifier that
is robust to $r\sim22$.  As a result, in addition to straight
colour-cuts, we also apply some colour restrictions on objects with
high $r$-band galaxy probability (referred to below as ``galprob'')
according to \nocite{sjd+02}{Scranton} {et~al.} (2002) in an attempt to remove contamination from
narrow emission line galaxies (NELGs; i.e. blue star-forming
galaxies) from our target list.

Bright sample objects are those with $18.0<g<21.15$ and that meet the
following conditions
\begin{eqnarray}
\begin{array}{rcclcccll}
{\rm A}) \;\, & u-g & < & 0.8 & \&\& & g-r & < & 0.6 & \&\& \\
 & r-i & < & 0.6 & & & & & \\
{\rm B}) \;\, & u-g & > & 0.6 & \&\& & g-i & > & 0.2 & \\
{\rm C}) \;\, & u-g & > & 0.45 & \&\& & g-i & > & 0.35 & \\
{\rm D}) \;\, & {\rm galprob} & > & 0.99 & \&\& & u-g & > & 0.2 & \&\& \\
 & g-r & > & 0.25 & \&\& & r-i & < & 0.3 & \\
{\rm E}) \;\, & {\rm galprob} & > & 0.99 & \&\& & u-g & > & 0.45. &
\end{array}
\end{eqnarray}
in the combination $A\&\&!B\&\&!C\&\&!D\&\&!E$, where cut A selects
UVX objects, cuts B and C eliminate faint F-stars whose metallicity
and errors push them blueward into the quasar regime, and cuts D and E
remove NELGs that appear extended in the $r$ band.  Among the bright
sample objects, those with $g>20.5$ were given priority in terms of
fibre assignment.

Faint sample objects are those with $21.15\le g<21.85$ and that meet the
following conditions
\begin{eqnarray}
\begin{array}{rcclcccll}
{\rm A}) \;\, & u-g & < & 0.8 & \&\& & g-r & < & 0.5 & \&\& \\
 & r-i & < & 0.6 & & & & & \\
{\rm B}) \;\, & u-g & > & 0.5 & \&\& & g-i & > & 0.15 \\
{\rm C}) \;\, & u-g & > & 0.4 & \&\& & g-i & > & 0.3 \\
{\rm D}) \;\, & u-g & > & 0.2 & \&\& & g-i & > & 0.45 \\
{\rm E}) \;\, & {\rm galprob} & > & 0.99 & \&\& & g-r & > & 0.3.
\end{array}
\end{eqnarray}
in the combination $A\&\&!B\&\&!C\&\&!D\&\&!E$, where cut A selects
UVX objects, cuts B, C and D eliminate faint F-stars whose
metallicity and errors push them blueward into the quasar regime, and
cut E removes NELGs.  These faint cuts are more restrictive than the
bright cuts to avoid significant contamination from main sequence
stars that will enter the sample as a result of larger errors at
fainter magnitudes.



\begin{figure}
\includegraphics[width=84mm]{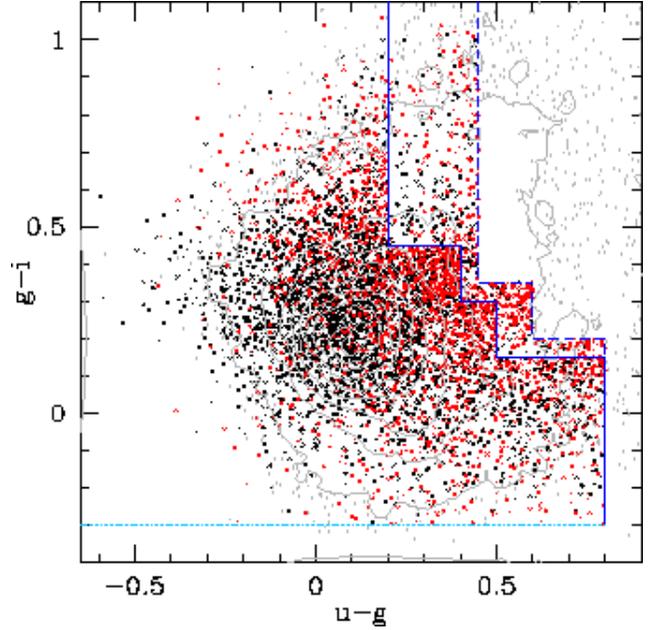}
\caption{Colours of spectroscopically identified 2SLAQ targets.
Confirmed quasars are shown in {\em black}, non-quasars in {\em red}.
For reference the colour distribution of $z<3$ SDSS-DR1 quasars
(Schneider et al.\ 2003)
are shown as faint grey contours/points.  The dashed
and solid dark blue lines show the $u-g$ and $g-i$ colour cuts for the
bright and faint samples, respectively.  The dotted cyan line shows
the boundary of the $g-i$ cut used to reject white dwarfs.}
\label{fig:fig1}
\end{figure}

Figure~\ref{fig:fig1} shows the $u-g$ vs.\ $g-i$ colour distribution
of objects satisfying these criteria for which we obtained new
spectra.  Objects confirmed to be quasars are shown in {\em black},
while those that are not quasars (mostly stars and NELGs) are shown in
{\em red}.  The locus of $z<3$ quasars from SDSS-DR1 \nocite{sfh+03}({Schneider} {et~al.} 2003) is
given by grey contours and points.  Solid blue, dashed blue and dotted
cyan lines show the faint sample colour cuts, the bright sample colour
cuts and the white dwarf cut, respectively.

\subsection{Sky location of imaging data}
\label{sec:skyloc}

\subsubsection{2003A and 2004A}

\begin{figure}
\includegraphics[width=84mm]{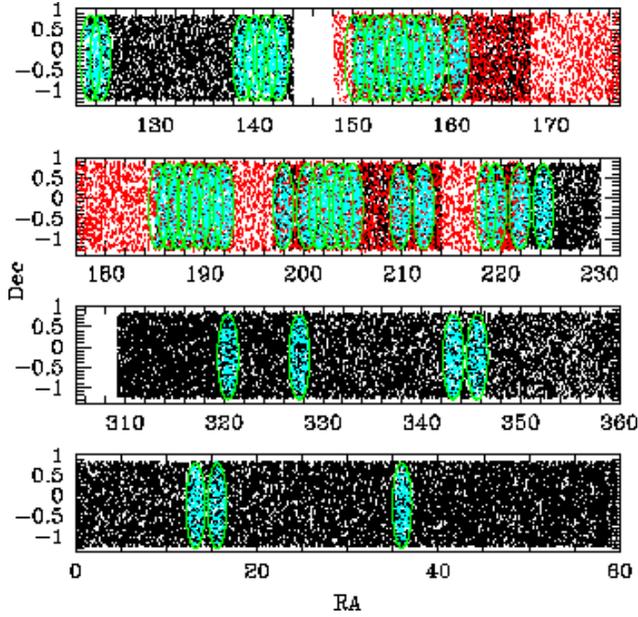}
\caption{RA and Dec distribution of 2dF targets selected from SDSS
imaging data ({\em black}) and previous 2dF observations ({\em red}).
2SLAQ observations are given by cyan points within 41 plates ({\em
green circles}).  The top two panels are semester A targets; the
bottom two panels are semester B targets.  Note the distortion of the
coordinate system; the spectroscopic plates are actually circular.}
\label{fig:fig2}
\end{figure}

For the first semester both of 2003 and 2004, we used the SDSS imaging
data (rerun 20; \nocite{slb+02,aaa+03}{Stoughton} {et~al.} 2002; {Abazajian} {et~al.} 2003) in the SDSS northern
equatorial scan (stripe 10) from runs 752, 756, 1239 and 2141; see
\nocite{yaa+00}{York} {et~al.} (2000) and \nocite{slb+02}{Stoughton} {et~al.} (2002) for the definition of the relevant
SDSS technical terms.  Run 756 was used for the northern part of the
stripe, while a combination of the other three runs was used for the
southern part of the stripe in an attempt to use the best quality data
(typically that with the best seeing).  The area of sky sampled was
further limited to regions where the SDSS image quality was deemed to
be good enough to use for targeting faint objects for spectroscopy,
specifically seeing $\le1\farcs8$ and $r$-band Galactic extinction
$\le0.2$ \nocite{sfd98}({Schlegel} {et~al.} 1998).  We also excluded any objects from SDSS camera
column 6, since 2dF cannot cover the full 2.5 degree wide SDSS stripe
and column 6 has the lowest quality data of all the columns as a
result of (relatively) poorer image quality at the edge of the camera
in these early SDSS data.  The final RA ranges were
$123\arcdeg<\alpha_{\rm J2000}<144\arcdeg$, $150\arcdeg<\alpha_{\rm
J2000}<168\arcdeg$, $185\arcdeg<\alpha_{\rm J2000}<193\arcdeg$,
$197\arcdeg<\alpha_{\rm J2000}<214\arcdeg$ and
$218\arcdeg<\alpha_{\rm J2000}<230\arcdeg$.  Whenever possible, we
tried to overlap areas with existing 2QZ spectroscopy to limit the
number of objects with $b_J<20.85$ that needed spectroscopic
confirmation.  SDSS spectroscopy limits the need for new $i<19.1$
spectra.  The two top panels of Figure~\ref{fig:fig2} illustrate the
area covered by our Semester A targets ($-1.259\arcdeg\le\delta_{\rm
J2000}\le0.840\arcdeg$).

\subsubsection{2003B}

For the second semester, our samples were limited to the following
combination of data (run, rerun, strip, $\alpha_{\rm J2000}$ range):
(2659, 40, 82N, $309.20\arcdeg<\alpha_{\rm J2000}<320.34\arcdeg$),
(2662, 40, 82N, $320.34\arcdeg<\alpha_{\rm J2000}<15.08\arcdeg$),
(2738, 40, 82N, $15.08\arcdeg<\alpha_{\rm J2000}<59.70\arcdeg$),
(2583, 40, 82S, $309.20\arcdeg<\alpha_{\rm J2000}<341.08\arcdeg$),
(3388, 40, 82S, $341.08\arcdeg<\alpha_{\rm J2000}<345.44\arcdeg$),
(3325, 41, 82S, $345.44\arcdeg<\alpha_{\rm J2000}<59.70\arcdeg$).
These reruns (40 and 41) represent post-DR1 data processing, which
includes a newer version of the photometric pipeline and improved
photometric calibration.  Again, camera column 6 was excluded and
these are all equatorial scans.  Note that there are no 2QZ
observations in this range.  The two bottom panels of
Figure~\ref{fig:fig2} illustrate the area covered by our Semester B
targets.

\subsubsection{Sky area}

The area of sky covered by our catalogue of targets for 2003/4A was
159.4 deg$^2$ with 20228 targets and for 2003B it was 230.2 deg$^2$
with 33160 total targets.  Thus we have a total area of 389.6 deg$^2$
and 53388 targets.  Of this area, this paper concentrates on only
those regions where we have obtained new spectra (see
Figure~\ref{fig:fig2}).  In semester 2003/4A, 34 plates were observed,
covering an area of 80.82 deg$^2$ -- as determined by the fraction of
targets within the plate areas (11075 of 53388).  In the second
semester, seven plates were observed, covering an area of 24.9 deg$^2$
(3407 of 53388 targets within the new plate areas).  Note that the
plates overlapped in 2003/4A, but not in 2003B.  The theoretical area
for 2003B given a plate radius of 1.05 deg is $24.3$ square degrees,
which compared to the area estimated by fraction of targets (24.9
deg$^2$) suggests that our estimate of the area has a roughly 2.5 per cent
error.  Thus the area covered by new plate observations is
$105.7\pm2.6$ deg$^2$.  Within these plate centers there are 14482
targets, of which 9120 have spectroscopic identifications, and among
those are 5645 quasars.

\section{Spectroscopic observations}

\subsection{The 2dF facility}

The Two Degree Field (2dF; \nocite{lct+02}{Lewis} {et~al.} 2002) facility at the
Anglo-Australian Telescope is a fibre fed multi-object spectrograph
and robotic fibre positioner.  The fibres are $140\mu$m in diameter,
which is roughly $2\farcs16$ at the centre of the plate and
$1\farcs99$ at the edges.  Two independent spectrographs use Tektronix
1024$\times$1024 CCDs with a range of diffraction gratings offering
resolutions between 10\AA\ and 2.2\AA\ over the optical wavelength
range.  During standard operation, 400 fibres are available for
simultaneous observation (200 per spectrograph) over a 2 degree
diameter field of view.  
The system is equipped with an atmospheric dispersion compensator
which enables 2dF observations to be taken over a wide wavelength
range, by ensuring that all wavelengths from the UV to NIR enter the
fibres.  However, differential spatial atmospheric refraction distorts
the field geometry and limits observations of equatorial fields to
$\pm1$ hour on either side of transit.

\subsection{2dF field configurations}

The 2SLAQ survey regions are centred close to the equator and are 2
degrees wide in declination.  To achieve optimal sky coverage while
still retaining a largely contiguous area, the 2dF field centres are
placed along the central declination of the two strips; $\delta_{\rm
J2000} = -00^{\circ} 12\arcmin 35\arcsec$ for the North Galactic Pole
(NGP) strip and $\delta_{\rm J2000} = -00^{\circ} 15\arcmin 00\arcsec$
for the South Galactic Pole strip; see \S~\ref{sec:skyloc}.  Each
field centre is separated by 1.2 degrees, although some early
observations of the NGP at the start of 2003 had field spacings of 1
degree.

The target list generated from the process described in
\S~\ref{sec:sample} is then merged with a target list of LRGs selected
from the same photometric data set \nocite{can+05}({Cannon et al.} 2005).  The sub-samples
within this combined data set are assigned different priorities which
determine the likelihood of a fibre being assigned to them in the 2dF
configuration process.  The priority values given to each sample are
listed in Table~\ref{tab:priorities}, where 9 is the highest and 1 the
lowest priority.  All available high priority targets are allocated
before moving to the next priority level.  For source densities much
greater than 400 per 2dF field, the 2dF configuration algorithm will
tend to give a non-uniform distribution of fibres allocated to objects
\nocite{can+05}({Cannon et al.} 2005).  Therefore the main samples of each of the LRG and QSO
data sets were randomly sampled to a surface density of 200 per 2dF
field and these given priority 8 and 6 for the LRGs and QSOs
respectively.  The remaining sources from these main samples were
given lower priority (7 and 5 respectively).  Other sub-samples, such
as bright QSOs and high-redshift candidates, were given lower
priority.

For the QSO sample we used the low resolution 300B grating (as used
for the 2QZ), but the LRG observations required the use of the higher
resolution 600V grating.  Therefore, one of the 2dF spectrographs is
configured with a 300B grating (spectrograph 1) while the second
(spectrograph 2) is configured with the 600V grating.  On each 2dF
field plate of 400 fibres, each block of 10 fibres (a retractor block)
goes to an alternate spectrograph, so that 200 fibres on each plate
are available for the QSOs and 200 for the LRGs.  2dF fibres are
limited to a maximum off-radial angle of 14 degrees, therefore there
are 20 small triangles surrounding the edge of the 2dF fields that are
inaccessible to the QSO spectrograph covering a total area of 0.43
deg$^2$.  The angular completeness function defined by this complex
field pattern is not relevant to the QLF analysis below, but it is
critical to accurate measurements of clustering.  Of the 200 fibres
available for the QSOs, 20 were allocated to positions known to be
blank sky to be used for sky subtraction.

\begin{table}
\begin{center}
\caption{2dF configuration priorities.}\label{tab:priorities}
\begin{tabular}{lc}
\hline
Sample & Priority \\
\hline
Guide stars     & 9 \\
LRG (main) random      & 8 \\
LRG (main) remainder     & 7 \\
QSOs ($g>20.5$) random & 6 \\
QSOs ($g>20.5$) remainder & 5 \\
LRG(extras)+hi-z QSOs & 4 \\
QSOs ($g<20.5$) & 3 \\
previously observed & 1 \\
\hline
\end{tabular}
\end{center}
\end{table}

\subsection{2dF observations and data processing}

Observations started in March 2003.  Each 2dF field was observed for a
minimum of four hours (more if weather was poor).  These four hours
were split over two nights to minimize the effects of changing
atmospheric refraction.  The 300B grating used gives a dispersion of
4.3\AA\ pixel$^{-1}$ and an instrumental resolution of 9\AA.  The
spectra cover the range 3700-7900\AA.  The data were reduced in real
time using the standard {\sc 2dFdr} pipeline (Bailey et al.\ 2003, MNRAS
submitted).  The exposure times increased if the conditions meant that
a pre-defined completeness limit (80 per cent) was not met.  Any
source that has a high S/N spectrum and a high-confidence
identification after the first night of observation has its fibre
assigned to previously unallocated sources for future observations of
the field.

The identification of QSOs and measurement of redshifts was done using
the {\sc AUTOZ} code that was developed for the 2QZ (see
\nocite{csb+01}{Croom} {et~al.} (2001, \S~3.1) and \nocite{csb+04}{Croom} {et~al.} (2004, \S~2.3.1) for details).
All spectra are then checked by eye to confirm the identifications.
Since spectroscopic processing is the same as that used for 2QZ
spectra (e.g., quasars must have broad [$>1000\,{\rm km\,s^{-1}}$]
emission lines), we treat 2SLAQ selected objects with 2QZ spectra as
if they were observed as part of the 2SLAQ programme.

\section{Completeness corrections}

In this section we explore and quantify the various effects that will
bias the quasar number counts and luminosity function.  In particular,
we address the photometric, coverage, spectroscopic and cosmetic
defect incompleteness of our sample.  In addition, we investigate the
difference between $g$ and $b_J$ magnitudes, Eddington bias,
morphology bias and the effects of variability.

\subsection{Coverage and spectroscopic completeness}

\begin{figure}
\includegraphics[width=84mm]{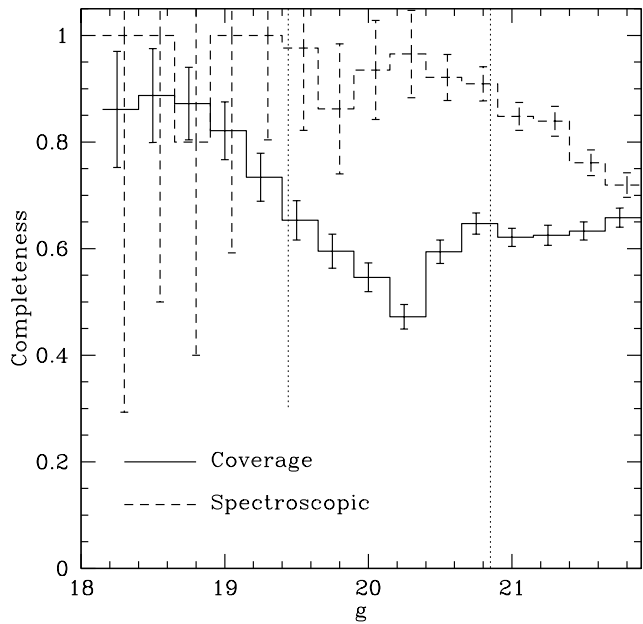}
\caption{Spectroscopic ({\em dashed line}) and coverage ({\em solid
line}) completeness fractions as a function of magnitude in 0.25 mag
bins.  These corrections were applied (in conjunction with the
photometric completeness correction [Fig.~\ref{fig:fig4}]) to
determine the corrected number counts.  The dotted vertical lines show
the boundaries of the SDSS ($i=19.1$, $g\sim19.32$) and 2QZ surveys
($b_J\sim g=20.85$).  The discontinuity in coverage completeness at
$g\sim20.5$ is caused by a prioritization of targets fainter than this
limit; see Table~1.}
\label{fig:fig3}
\end{figure}

We have not obtained spectra of all our quasar candidates in the 105.7
deg$^2$ analyzed in this paper.  Thus we must compute the ``coverage''
completeness of our sample, which multiplied by the area yields the
effective area of the survey.  Since we are combining data from three
distinct surveys (SDSS, 2QZ, and 2SLAQ) in order to increase our
dynamic range, it is necessary to compute this correction as a
function of magnitude.  The coverage completeness is computed under
the assumption that the fraction of objects that remain unobserved (at
a given magnitude) will be quasars at the same rate as those that are
observed.  This assumption is reasonable given that the objects
observed are chosen at random.  Figure~\ref{fig:fig3} shows the
coverage completeness ({\em solid line}) that we compute as a function
of magnitude.

In addition to the coverage completeness, we must correct for those
cases in which our spectroscopy does not yield an unambiguous
identification.  Assuming that the fraction of unidentified objects
will be quasars at the same rate as those among identified objects (as
a function of magnitude), we derive a spectroscopic incompleteness as
shown by the {\em dashed} line in Figure~\ref{fig:fig3}.  This
assumption arguably may tend to overestimate the number of
unidentified quasars since the spectroscopic completeness may
additionally be a function of redshift (because of emission line
effects which generally facilitate quasar identification) and that
{\em any} completeness determination is surely to be a lower-limit.
However, our spectroscopic completeness is generally high (70 per cent at the
faint limit, 90 per cent brighter), thus any second-order corrections will
have a minimal impact.  Furthermore, comparison with supplementary
identifications based solely on photometry and photometric redshifts
(\S~\ref{sec:qlf}) suggests that this assumption is reasonable.  In
practice we have treated the spectroscopic completeness as if the
unidentified objects simply had not been observed, which facilitates
the application of these corrections to our model of the luminosity
function.

\subsection{Photometric completeness}
\label{sec:photcomp}

The incompleteness of our sample due to colour cuts is a strong
function of both redshift and magnitude since the colours of quasars
change significantly with redshift and fainter quasars have larger
errors.  We quantify this incompleteness by running our selection
algorithm on a sample of simulated quasars that were designed to test
the SDSS's quasar target selection algorithm; see \nocite{fan99}{Fan} (1999) and
\nocite{ric+05}{Richards et al.} (2005).  The primary independent variable in the simulations
is the spectral index distribution, which is given by a Gaussian
distribution with $\alpha_{\nu}=-0.5\pm0.3\;(f_{\nu} \propto
\nu^{\alpha_{\nu}})$, which is in good agreement with the composite
SDSS quasar spectrum given by \nocite{vrb+01}{Vanden Berk} {et~al.} (2001).  Blueward of the
Ly$\alpha$ emission line we instead use a spectral index of
$\alpha_{\nu}=-1.5\pm0.17$, consistent with \nocite{tzk+02}{Telfer} {et~al.} (2002); this
spectral index is taken to be uncorrelated with the optical/UV
spectral index.  Only the spectral index, the Ly$\alpha$ equivalent
width and the Ly$\alpha$ forest strength vary; all other emission
lines are fixed relative to Ly$\alpha$.

\begin{figure}
\includegraphics[width=84mm]{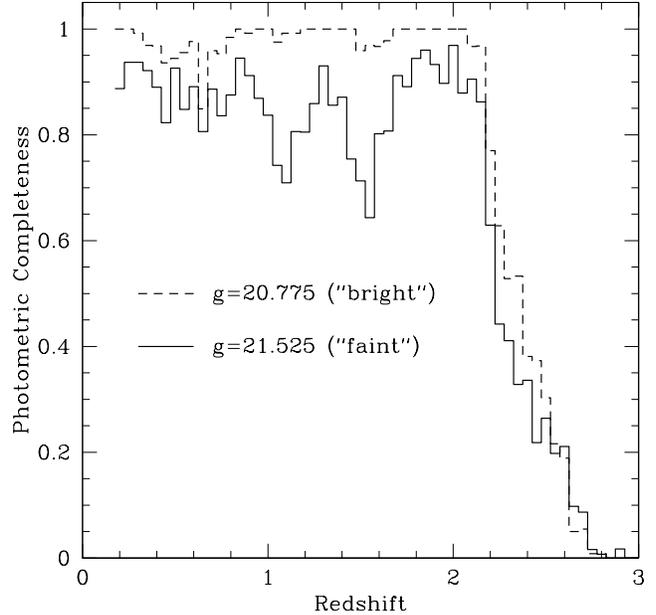}
\caption{Completeness as a function of redshift and $g$ magnitude
based on simulated quasars.  Representative magnitude ranges are shown
for ``bright'' and ``faint'' samples with redshift intervals of
0.05.}
\label{fig:fig4}
\end{figure}

Figure~\ref{fig:fig4} shows the selection completeness to these
simulated quasars as a function of redshift and $g$ magnitude.  Two
representative ranges are shown, with bins $0.25$ mag wide centered on
$g=20.775$ and 21.525.  The $g=20.775$ completeness curve ({\em dashed
line}) is representative of the ``bright'' sample, whereas the
$g=21.525$ curve ({\em solid line}) is representative of the
``faint'' sample (except for the faintest bin since it extends to
$g=21.9$ and the sample only goes to $g=21.85$).  Incomplete redshift
regions occur when photometric errors are large and/or
emission/absorption lines bring the colours of the quasars near/across
the colour cuts \nocite{rfs+01}(e.g. {Richards} {et~al.} 2001).

\subsection{Correction for cosmetic defects}

Certain cosmetic defects within the imaging data cause quasars to be
missed from our sample.  Thus, we need to make a correction for
cosmetic defects in the SDSS data, specifically for those objects that
fail the fatal/non-fatal error tests \nocite{rfn+02}({Richards} {et~al.} 2002).  One way to
quantify this is to assume that any cosmetic defects that prevent the
selection of a particular quasar in the SDSS imaging are unlikely to
have been present in the 2QZ imaging inputs.  With the exception of
blended objects, this assumption should be roughly true.  Thus we
match the NGP sample of quasars from the 2QZ to our initial catalogue
of Semester A targets (with only the fatal and non-fatal errors,
$i<22$ and $\sigma_i<0.2$ cuts applied).  Since the 2QZ only went to
$b_J=20.85$, the $i$ magnitude cut should not cause us to lose many
quasars; however, the fatal and non-fatal error cuts (i.e. cosmetic
defects) {\em will} cause quasars to be lost.  The fraction of 2dF
quasars that are not among our initial SDSS-imaging selected sample
gives us an estimate of the fraction of quasars that are missed due to
cosmetic defects.  We find that this fraction is $\sim5$ per cent.  A
similar fraction is derived by \nocite{van+05}{Vanden Berk et al.} (2005) based on an empirical
analysis of the point-source completeness of the SDSS quasar
catalogue.  We apply this correction independent of
magnitude\footnote{But note that \nocite{van+05}{Vanden Berk et al.} (2005) find that this
completeness is a function of magnitude; however, the completeness has
not been determined at the faint limits to which we are probing, so we
assume a uniform value.} and redshift in addition to the coverage,
spectroscopic and photometric completeness corrections described
above.  Losses due to blending of sources will increase this
completeness correction; for our purposes such losses are assumed to
be smaller than the other corrections that we apply.

\subsection{Eddington bias}


\begin{figure*}
\includegraphics[width=168mm]{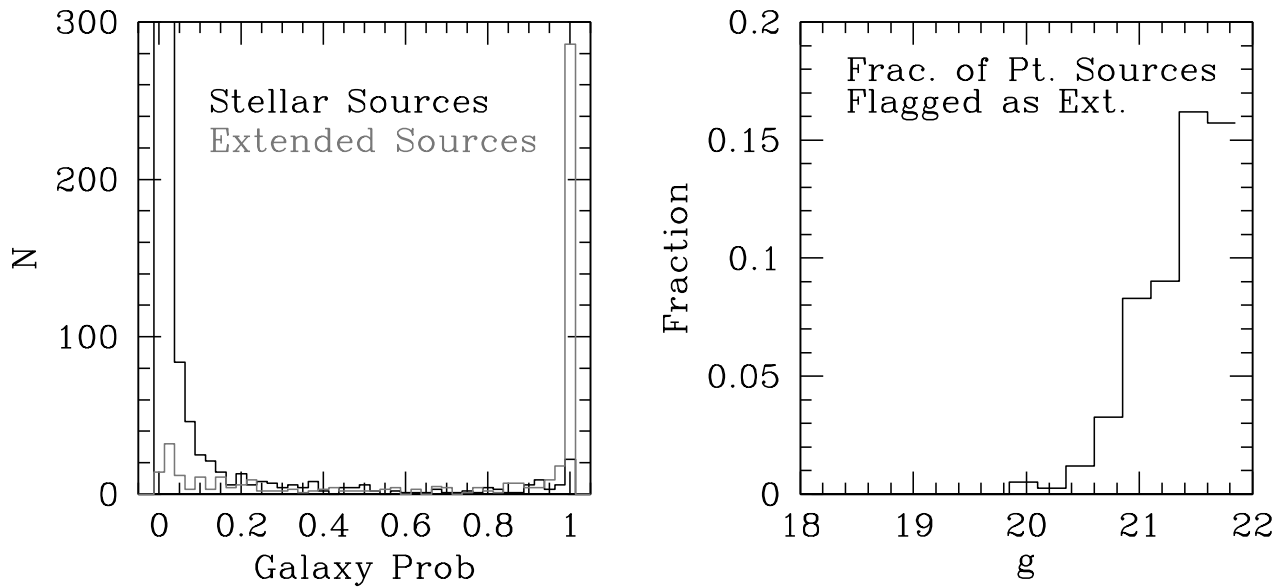}
\caption{{\em Left:} Number of spectroscopically confirmed quasars
classified by PHOTO as stellar ({\em black}) and as extended ({\em
grey}) as a function of Bayesian galaxy probability 
(Scranton et al.\ 2002).
{\em Right:} Fraction of point-like quasars (as determined by the
Bayesian analysis) that are mis-classified by PHOTO as extended -- as
a function of magnitude.}
\label{fig:fig5}
\end{figure*}

Eddington bias is the distortion of the object number counts as a
function of magnitude that occurs when photometric uncertainty causes
errors in distributing sources into their proper magnitude bins.  The
relationship between the observed and actual differential number
counts, $A(m)$, is given by \nocite{pet97}{Peterson} (1997):
\begin{equation}
A(m) \approx A_{obs}(m)\left[1 - \frac{1}{2}\left(\frac{\sigma\kappa}{\log e}\right)^2\right],
\end{equation}
where $\sigma$ is the Gaussian error in the magnitude, $\kappa$ is
defined by the integrated number counts relation $N(m) \propto
C10^{\kappa m}$, and $\log A(m) = C + \kappa m$.  If the product of
the slope and the error ($\sigma\kappa$) increases with magnitude then
the observed slope is steeper than the intrinsic slope; for decreasing
$\sigma\kappa$ the observed slope is flatter than the intrinsic slope.
For our sample the correction term in brackets above is $\ge0.98$ for
all magnitude bins, thus we have applied no correction for Eddington
bias.

\subsection{Morphology bias}

Our sample includes objects that the SDSS photometric pipeline ({\sc PHOTO};
\nocite{lgi+01}{Lupton} {et~al.} 2001) classifies as extended.  The rationale for this
decision is summarized in Figure~\ref{fig:fig5} which shows that at
the faintest limits of our survey, a significant fraction of point
sources are mis-classified by the photometric pipeline as extended
(assuming that the Bayesian analysis of \nocite{sjd+02}{Scranton} {et~al.} 2002 represents
ground truth).  The right-hand panel shows that this is a function of
magnitude.  The left-hand panel shows the Bayesian galaxy probability
distribution for both point-like (stellar) and extended quasars as
classified by {\sc PHOTO}.

The inclusion of extended sources can lead to a bias.  Specifically,
since many of the Semester A targets have been observed as part of the
2QZ and since the 2QZ did not target extended sources, our new
observations will be preferentially biased towards extended sources.
Thus our corrections from the number of objects observed to the number
of objects targeted may be skewed since it assumes that new
observations will yield quasars at the same rate as old observations.
However, we find that, although the contamination among extended
sources is larger than for point sources, the {\em shape} of the
corrections as a function of magnitude are not significantly different
and thus our analysis of the shape of the QLF should not be adversely
affected.

\subsection{$g$ vs. $b_J$}

\begin{figure}
\includegraphics[width=84mm]{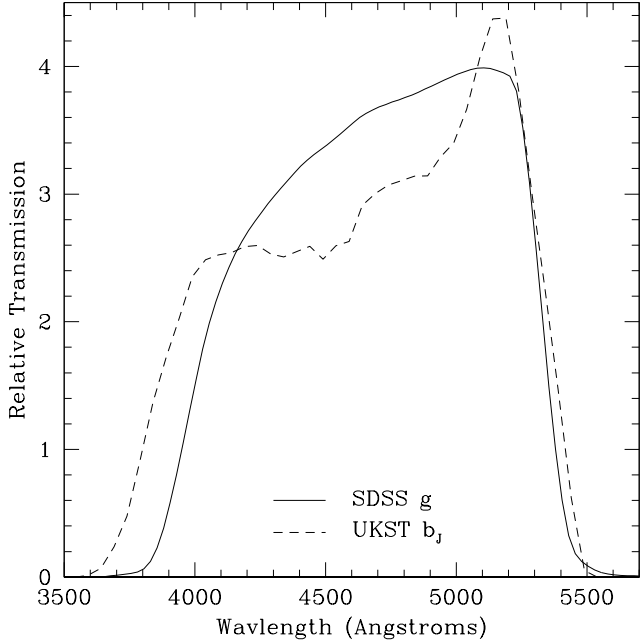}
\caption{UKST $b_J$ (P. Hewett 2004, private communication) and SDSS
$g$ transmission curves for 1 airmass and normalized such that $\int
S_{\lambda} \frac{d\lambda}{\lambda} = 1$.  Both curves are given in
terms of detector quantum efficiency, which means the $g$ curve is
shown as published, but the $b_J$ curve has been multiplied by
wavelength.}
\label{fig:fig6}
\end{figure}

\begin{figure}
\includegraphics[width=84mm]{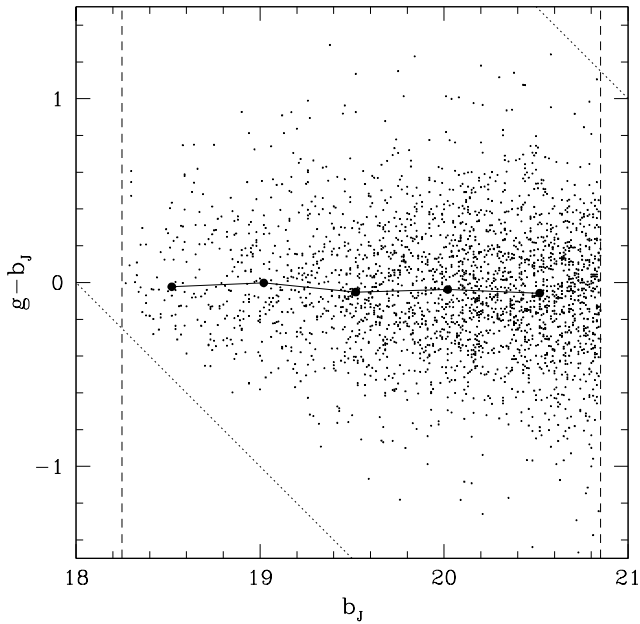}
\caption{$g-b_J$ magnitude differences versus $b_J$, where $g$ is from
the SDSS and $b_J$ is from the 2QZ.  The dotted lines show the $g$
magnitude limits of the 2SLAQ sample.  The dashed lines show the $b_J$
magnitude limits of the 2QZ survey.  The points connected by a solid
line give the median $g-b_J$ as a function of $b_J$ in 0.5 mag bins.
The median over all $b_J$ is $\langle\!g-b_J\!\rangle\;=-0.045$.
Neither $g$ or $b_J$ are extinction corrected in this
plot.}
\label{fig:fig7}
\end{figure}

To properly compare our 2SLAQ results to those of the 2QZ, we
determine the relationship between the SDSS $g$ band and the $b_J$
band used by 2dF.  Figure~\ref{fig:fig6} shows the two transmission
curves, which are quite similar.  The $b_J$ curve was kindly provided
by Paul Hewett (2004, priv. comm.).  The $g$ curve is as
taken from the SDSS web
site\footnote{http://www.sdss.org/dr3/instruments/imager/filters/g.dat}
-- except that it has been converted from 1.3 to 1.0 airmasses (to
match the $b_J$ curve).  In Figure~\ref{fig:fig7}, we plot the $g-b_J$
magnitude difference versus $b_J$ for all of the 2QZ quasars in our
sample.  This plot shows that even considering the scatter in $g-b_J$,
the $g$ band magnitude limits of our current sample completely
encompass the 2QZ quasars.

To convert $b_J$ to $g$ we simply compute the median $g-b_J$
difference, which is shown as a function of $b_J$ by points connected
by solid lines in Figure~\ref{fig:fig7}.  The median for the whole
sample is $\langle\!g-b_J\!\rangle\;=-0.045$, with no significant
dependence on $b_J$.  Given the empirical similarity of the $g$ and
$b_J$ magnitudes, and that the error in the computed median is of
order the median itself, we have simply taken $b_J$ as an exact
surrogate for $g$ in our comparison of the number counts and
luminosity functions.

Much of the scatter between $b_J$ and $g$ is caused by variability in
the $\ge$20 years between the epochs when the $b_J$ and $g$ data were
taken -- in contrast with the simultaneity of the SDSS 5-band imaging
data.  The scatter in $g-b_J$ is $\sigma_{g-b_J}=0.25$ at $b_J=18.475$
and $\sigma_{g-b_J}=0.38$ at $b_J=20.725$.  At least $0.15$ mag of
this error is due to photometric error in $b_J$
\nocite{csb+04}({Croom} {et~al.} 2004, Fig.~9); roughly $0.02$ and
$0.035$ is due to photometric error in $g$.  Thus, most of the scatter
(roughly $\sigma\sim0.2$) is thus caused by variability.  Variability
adds uncertainty to the magnitude distribution in the same manner as
photometric errors and thus can modify the number counts through
Eddington bias.  Proper treatment of variability in light of quasar
number counts is complicated, ideally using long terms averages of the
quasars under consideration.  However, we can estimate the effect that
variability has on the slope of the number counts.  If the variability
amplitude is constant with magnitude, then variability will cause a
slight flattening of the observed distribution due to the number
counts being steeper at the bright end than at the faint end.  For
$\sigma_{\rm var}=0.2$, at $g\sim18.5$ the number counts will be
over-estimated by $\sim$8\% and at $g\sim20.7$ they will be
over-estimated by $\sim$1\%, which produces a negligible ($\sim$2\%)
change in slope over this range.

\section{Number counts and luminosity function}
\label{sec:qlf}

\subsection{Redshift and absolute magnitude distributions}

\begin{figure}
\includegraphics[width=84mm]{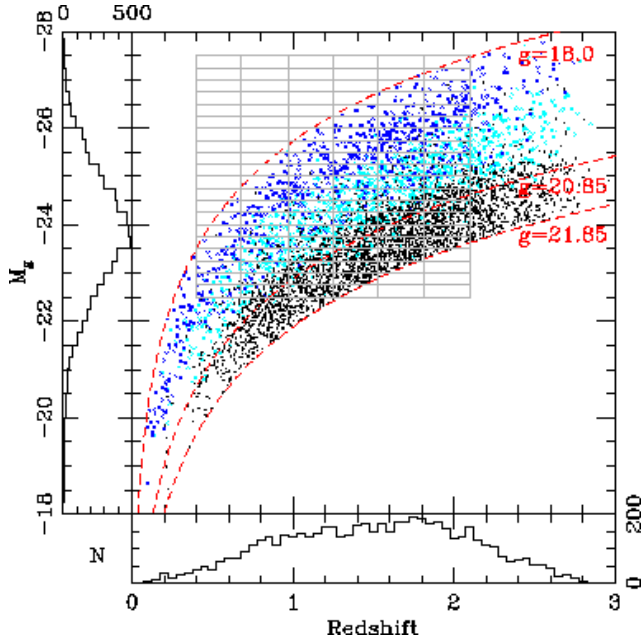}
\caption{Absolute $g$ magnitude versus redshift for all confirmed
quasars in our sample.  Blue crosses are SDSS quasars, cyan crosses
are 2QZ quasars, while black dots represents quasars confirmed by
2SLAQ.  The bottom and side histograms show the 1D distributions of
$M_g$ and redshift.  The dashed red lines show the bright and faint
magnitude limits of this survey ($g=18.0$ and $g=21.85$) and the
approximate limit of the 2QZ survey in $g$ ($b_J=20.85$).  The grid of
grey lines outline the bins used for determining the quasar luminosity
function.}
\label{fig:fig8}
\end{figure}

Having discussed the various completeness corrections, we can now
determine the number counts and luminosity function of our sample.
Figure~\ref{fig:fig8} shows the $M_g$ vs.\ redshift distribution of
spectroscopically confirmed 2SLAQ, 2QZ and SDSS quasars in our sample
-- within the boundaries of new plate observations (105.7 deg$^2$).
The absolute $g$ magnitude, $M_g$, is computed using luminosity
distances in the cosmology given in \S~1 according to the prescription
of \nocite{hog99}{Hogg} (1999) and with the (albeit poor, but commonly
used) assumption of a universal power-law continuum of
$\alpha_{\nu}=-0.5$ ($f_{\nu}\propto\nu^{\alpha}$).\footnote{Ideally
we would determine a spectral index for each individual object.
However, this requires better spectrophotometry/photometry at the
faint end than 2SLAQ provides.  Fortunately, the errors induced by
assuming a fixed spectral index are mitigated by the $z<2.1$ nature of
our analysis (the errors increase with redshift) and the fact that the
majority of quasars have roughly this spectral index.}  Black, cyan
and blue points represent new 2SLAQ quasars, previously confirmed 2QZ
quasars and previously confirmed SDSS quasars, respectively.  Dashed
red lines at $g=18.0$ and 21.85 demarcate the $g$ magnitude boundaries
of our sample.  In addition we show the $g \sim b_J = 20.85$ limit of
the 2QZ survey.  The histograms to the left and bottom of the figure
show the one-dimensional distribution of sources in $M_g$ and
redshift.  We further overlay a grid which highlights the magnitude
and redshift bins that were used in the construction of the
\nocite{csb+04}{Croom} {et~al.} (2004) QLF and will also be used for
determining the binned 2SLAQ QLF.

\subsection{Number counts}

\begin{figure}
\includegraphics[width=84mm]{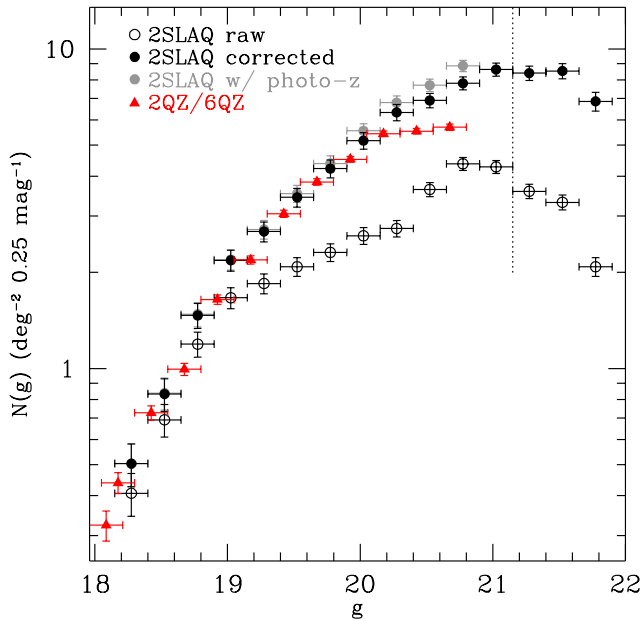}
\caption{2SLAQ number counts ({\em black circles}) compared to 2QZ
number counts ({\em red filled triangles}) from Croom et al.\
(2004). The 2SLAQ number counts are given both as raw (i.e. observed;
{\em open circles}) and corrected ({\em filled circles}) number
counts.  2SLAQ number counts are also given (as grey circles) after
including photometrically identified quasars (with photometric
redshifts) from
Richards et al.\ (2004).  Quasars are restricted to those with
$M_g<-22.5$ and $0.4<z<2.1$ (3889 quasars) for comparison with the 2QZ
number counts.  The dotted vertical line marks the dividing line between
the 2SLAQ bright and faint samples.}
\label{fig:fig9}
\end{figure}

Figure~\ref{fig:fig9} shows the differential number counts as a
function of $g$ magnitude in bins of 0.25 mag, both corrected ({\em
solid circles}) and uncorrected ({\em open circles}) for the various
sources of incompleteness (error bars are Poisson).  Number counts
from 2QZ \nocite{csb+04}({Croom} {et~al.} 2004) are shown in red for comparison.  This diagram
only includes quasars with $M_g<-22.5$ and $0.4<z<2.1$.  

From Figure~\ref{fig:fig9} we see that to $g\sim20.2$ the agreement
between 2SLAQ and 2QZ is quite good, but there is a discrepancy
between the two studies at the faint end: 2SLAQ suggesting a higher
density of faint objects than 2QZ.  We note that the shape of the
distribution is clearly better fit by a double power law than a single
power law (demonstrating the turnover in the distribution towards
fainter quasars), but that the change in slope is more subtle than the
distinctive ``break'' near $g\sim19.5$ that is sometimes found in such
analyses \nocite{bfsp87}(e.g. {Boyle} {et~al.} 1987).  This behaviour is qualitatively
consistent with that found by \nocite{wwb+03}{Wolf} {et~al.} (2003) from the COMBO-17 survey
and is inconsistent with the single power-law form found in variability
selected samples \nocite{hv95}(e.g. {Hawkins} \& {Veron} 1995, but see Ivezi\'c et al.\ 2004).

We have shown (as open circles) the raw number counts to give the
reader an idea of the absolute lower limits on the points and the size
of the completeness corrections that have been applied.  The coverage
corrections are straightforward and should be fairly robust (perhaps
less so in the 3 faintest bins due to the more restrictive selection
criteria and larger photometric error).  In fact, we could have simply
corrected the effective area as a function of magnitude and shown the
(more complete and much smoother) area-corrected raw counts.  However,
as we are splicing together three samples (SDSS, 2QZ, and 2SLAQ) to
provide spectroscopic coverage of our targets, it seems appropriate to
fully disclose the magnitude dependence of the coverage completeness
within the 105.7 deg$^2$ area covered by the 2SLAQ plates.  As a check
on our correction terms, we have also matched our unobserved and
unidentified objects to the photometric quasar candidate catalogue of
\nocite{rng+04}{Richards} {et~al.} (2004a), in attempt to ``observe''
a larger fraction of our quasar candidates (to $g<21$).  The objects
from \nocite{rng+04}{Richards} {et~al.} (2004a) are expected to be 95
per cent accurate (averaged over all magnitdues) with respect to
quasar classification, with 90 per cent having photometric redshifts
correct to $|\Delta z|\pm0.3$ for the redshift range considered here
\nocite{wrs+04}({Weinstein} {et~al.}  2004).  The result of including
these photometric identifications is shown by the grey points in
Figure~\ref{fig:fig9} and lends credence to the steeper faint-end
number counts relations that we derive solely from our
(completeness-corrected) spectroscopic sample.  This comparison is
meant purely as a sanity check.  The differences between the
spectroscopic (black cirles) and photometric (grey circles) number
counts are consistent with the expected decrease in efficiency of the
R04 photometric catalog with fainter magnitude, thus supporting the
accuracy of our completeness determinations (and our corrected number
counts).

\begin{figure}
\includegraphics[width=84mm]{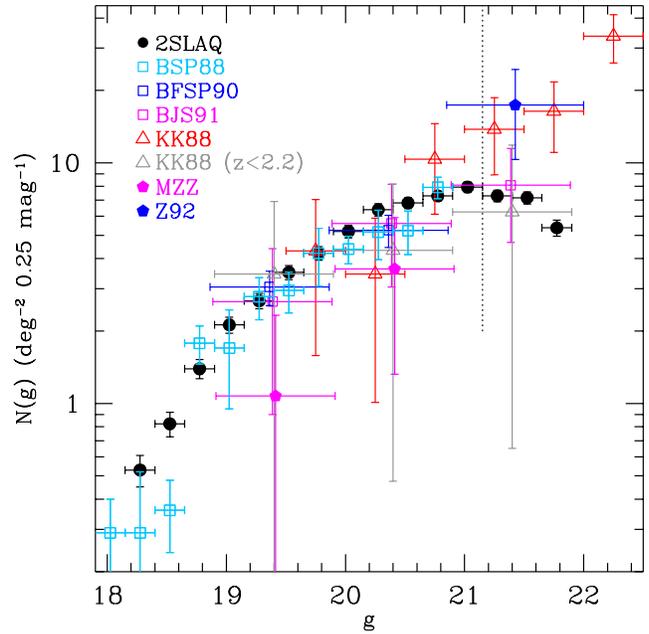}
\caption{Comparison of 2SLAQ quasar number counts to previous deep
samples.  2SLAQ quasars are limited to $M_g<-23$ and $0.6<z<2.2$ to
mimic the exclusion of extended sources (which mostly have $z<0.6$ or
$M_g>-23$).  Open squares indicate points from Boyle and
collaborators, specifically Boyle, Shanks \& Peterson (1988) [cyan];
Boyle, Fong, Shanks \& Peterson (1990) [blue]; Boyle, Jones \& Shanks
(1991; BSJ91) [magenta].  Open triangles refer to Koo \& Kron (1988),
where the red triangles are for $0.9<z<3.0$ and the grey triangles are
for $z<2.2$ (and $z>0.9$) as given by Table~8 in BJS91.  Filled
magenta pentagons refer to Marano, Zamorani \& Zitelli (1988), as
given by Table~8 in BJS91.  The filled blue pentagon is derived from
Zitelli et al. (1992).}
\label{fig:fig10}
\end{figure}

\begin{figure}
\includegraphics[width=84mm]{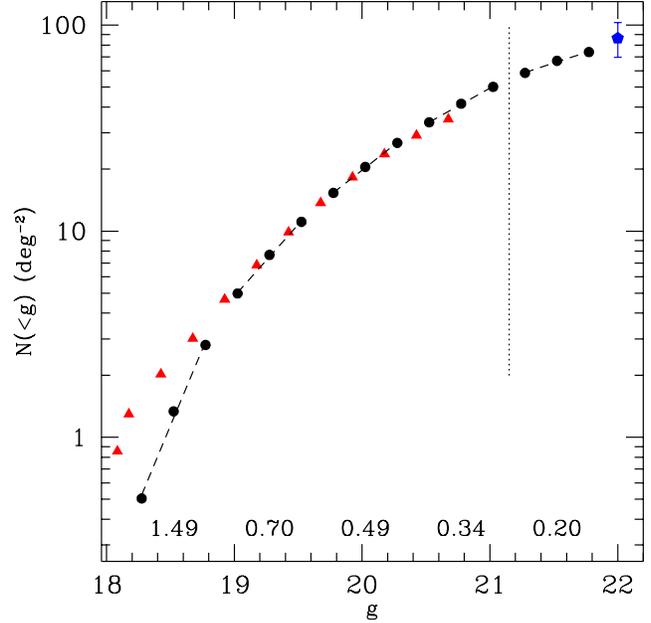}
\caption{Cumulative 2SLAQ ({\em black circles}) and 2QZ ({\em red
triangles}) as a function of $g$.  The numbers at the bottom indicate
the slope of the 3-bin least square fits shown by the series of dashed
lines.  Note that the bright limit of the 2SLAQ data is $g=18$ which
causes a deficiency in the cumulative number counts at the bright end;
at the faint end this lack of bright quasars makes little difference.
While there is no strong characteristic break, the number counts
clearly flatten with fainter magnitude.  The blue pentagon shows the
cumulative $J$-band number counts from Zitelli et al.\ (2002).}
\label{fig:fig11}
\end{figure}

We further compare our results to a number of other samples of faint
quasars that pre-date the 2QZ sample.  This comparison is shown in
Figure~\ref{fig:fig10}.  Here we have restricted our sample to
$0.6<z<2.2$ and $M_g<-23$ to best mimic the limits of these previous
surveys which generally excluded extended sources (which typically
have $z<0.6$ or $M_g>-23$).  We specifically compare our 2SLAQ results
to the samples of \nocite{bsp88}{Boyle} {et~al.} (1988), \nocite{kk88}{Koo} \& {Kron} (1988), \nocite{mzz88}{Marano}, {Zamorani} \& {Zitelli} (1988),
\nocite{bfsp90}{Boyle} {et~al.} (1990), \nocite{bjs91}{Boyle}, {Jones} \& {Shanks} (1991) and \nocite{z92}{Zitelli} {et~al.} (1992), where Table~8 in
\nocite{bjs91}{Boyle} {et~al.} (1991) is the source of the \nocite{bfsp90}{Boyle} {et~al.} (1990), \nocite{z92}{Zitelli} {et~al.} (1992), and
\nocite{kk88}{Koo} \& {Kron} (1988) [$z<2.2$] points.  The redshift ranges and magnitude
calibrations between all of these samples do not match exactly, but
they suffice to give the reader an idea of how our results compare
with past work.  In particular, in comparison with previous work we
note that while the 2SLAQ data show an excess at $20<g<20.6$, it
generally shows a deficit for $g>20.6$.  The one exception is the
faintest $z<2.2$ point from \nocite{kk88}{Koo} \& {Kron} (1988); however, that sample has a
lower redshift limit of $z\sim0.9$, whereas our sample extends to
lower redshift.  Overall, to the limit of our bright sample
($g<21.15$), our agreement with previous work is well within the
errors.  Fainter than $g=21.15$, if anything the 2SLAQ counts are
deficient, but are still consistent considering the large coverage and
spectroscopic completeness corrections at these limits.
Figure~\ref{fig:fig11} shows the cumulative 2SLAQ and 2QZ/6QZ quasar
number counts.  At the limit of the 2SLAQ survey, the cumulative
counts compare well with the $J=22$ cumulative counts ($86.3\pm16.5$)
from \nocite{z92}{Zitelli} {et~al.} (1992).  The slope of the cumulative counts are given as
3-bin averages by the dashed lines and the numbers at the bottom of
the plot.  The brightest 2SLAQ points are unreliable as 2SLAQ does not
include quasars brighter than $g=18$.  The cumulative 2QZ/6QZ number
counts gives a better idea of the slope at the bright end.
Table~\ref{tab:cum} shows a comparison of the cumulative number counts
predicted by the \nocite{bsc+00}{Boyle} {et~al.} (2000), \nocite{csb+04}{Croom} {et~al.} (2004) and 2SLAQ best fit
maximum likelihood parameterizations (assuming a double power-law and
luminosity evolution characterized by a 2nd order polynomial) for
$g>16.0$ and $0.3<z<2.2$.

\begin{table}
\begin{center}
\caption{Cumulative number count comparison.  The cumulative number
counts are shown for the Boyle et al.\ (2000), Croom et al.\ (2004)
and 2SLAQ maximum likelihood parameterizations for $16<g<mag$ and
$0.3<z<2.2$ in unit of counts per square degree.}\label{tab:cum}
\begin{tabular}{cccc}
\hline
$<mag$ & Boyle00 & Croom04  & 2SLAQ \\
\hline
20.0 & 15.87 & 17.50 & 18.96 \\
20.5 & 26.99 & 28.27 & 31.09 \\
21.0 & 41.68 & 40.22 & 47.79 \\
21.5 & 59.45 & 52.01 & 69.13 \\
22.0 & 78.88 & 62.46 & 93.77 \\  
\hline
\end{tabular}
\end{center}
\end{table}

\subsection{Luminosity function}

\begin{figure*}
\includegraphics[width=168mm]{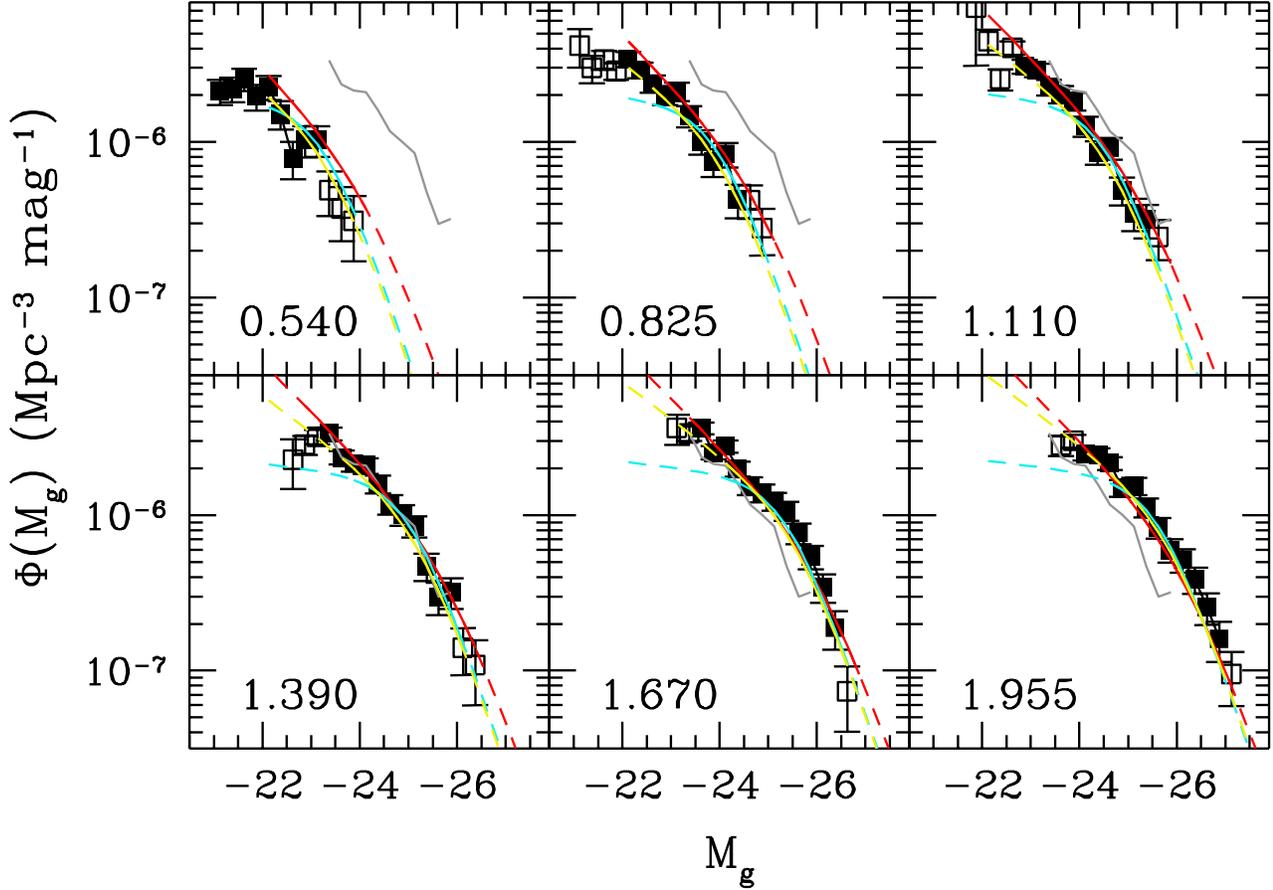}
\caption{Faint quasar luminosity function for 2SLAQ quasars.
Redshift bins are the same as for 
Croom et al.\ (2004).  Open points are incomplete bins (see
Fig.~\ref{fig:fig8}).  Corrections for photometric, coverage and
spectroscopic incompleteness have been applied.  Only the quasars
within the 41 new plate centers with $0.4<z<2.1$ are included.  The
$z=1.39$ data are repeated as grey curves in each panel for
comparison.  Best fit models (luminosity evolution characterized by a
2nd order polynomial in redshift) are shown from Boyle et al.\ (2000) [yellow],
Croom et al.\ 2004 [cyan] and this work (red) as solid and dashed
lines.  The dashed parts of the best fit lines indicate where the fits
have been extrapolated beyond the data.  The faint quasar data suggest
a steeper slope than the
Croom et al.\ (2004) models.}
\label{fig:fig12}
\end{figure*}

Figure~\ref{fig:fig12} shows two determinations of luminosity function
derived from our sample.  We first use the \nocite{pc00}{Page} \& {Carrera} (2000) implementation
of the $1/V$ method \nocite{sch68,ab80}({Schmidt} 1968; {Avni} \& {Bahcall} 1980), which is shown by the points
with error bars.  This implementation corrects for incompleteness at
both the bright and faint limits of the survey.  These incomplete bins
(those not filled in Figure~\ref{fig:fig8}) are shown as open rather
than closed points to indicate that they have been corrected for
partial coverage of the bin.  However, we note that the \nocite{pc00}{Page} \& {Carrera} (2000)
correction for incomplete bins is not fully accurate since the
(relatively large) $z-M_g$ bins are not uniformly sampled, see
Figure~\ref{fig:fig8}.  The redshifts are the same as those in
Figure~20 of \nocite{csb+04}{Croom} {et~al.} (2004) for ease of comparison. The size of the
redshifts bins is $\Delta z=0.283$ and the $z=1.39$ data are repeated
as grey lines in each panel.  

We next give the luminosity function as derived from a maximum
likelihood analysis; these are plotted as dashed/solid lines, the
dashed part indicating extrapolation beyond the data used for the fit.
The cyan lines show the best fit double power-law model (see below)
from row~1 in Table~6 of \nocite{csb+04}{Croom} {et~al.} (2004), which provide a poor fit at
the faint end.  The yellow lines show a similar model from row~1 in
Table~3 of \nocite{bsc+00}{Boyle} {et~al.} (2000) (corrected to our cosmology), which has a
steeper faint-end slope.  Our own fit is shown in red and was derived
as described below.

We have assumed a luminosity function in the standard form of a
double-power law \nocite{pet97,csb+04}({Peterson} 1997; {Croom} {et~al.} 2004)\footnote{We remind the reader of
the well-known sign error in \nocite{bsc+00}{Boyle} {et~al.} (2000) whereby (in the convention
used herein) the first equation in Section~3.2.2 of \nocite{bsc+00}{Boyle} {et~al.} (2000)
should have negative signs on $\alpha$ and $\beta$ and the entries for
$\alpha$ and $\beta$ in Tables~2 and 3 should be multiplied by $-1$.
In addition, equation~10 in \nocite{csb+04}{Croom} {et~al.} (2004) and the equivalent equation
in Section~3.2.2 of \nocite{bsc+00}{Boyle} {et~al.} (2000) are missing a $1/L^*$ factor in the
numerator.}
\begin{equation}
\Phi(L_g,z) = \frac{\Phi(L_g^*)/L_g^*}{(L_g/L_g^*)^{-\alpha} + (L_g/L_g^*)^{-\beta}},
\end{equation}
or
\begin{equation}
\Phi(M_g,z) = \frac{\Phi(M_g^*)}{10^{0.4(\alpha+1)(M_g-M_g^*)} + 10^{0.4(\beta+1)(M_g-M_g^*)}}.
\end{equation}
We assume that the evolution with redshift is characterized by pure
luminosity evolution (individual quasars getting fainter from $z=2$ to
today), with the dependence of the characteristic luminosity described
by a 2nd-order polynomial in redshift as in \nocite{csb+04}{Croom} {et~al.} (2004) where
\begin{equation}
M_g^*(z) = M_g^*(0) - 2.5(k_1z+k_2z^2).
\end{equation}
Note that this form assumes symmetric redshift evolution about the
peak.  This assumption is appropriate for UVX samples such as this
one, but will break down for samples that extend to higher redshifts
\nocite{ric+05}(e.g. {Richards et al.} 2005).

We compute the maximum likelihood solution via Powell's method
\nocite{ptv+92}({Press} {et~al.} 1992) using the form given by \nocite{fss+01}{Fan} {et~al.} (2001b).  We first
attempt to determine the best fit parameters by allowing all of the
parameters to vary.  The resulting parameters are given in the last
row of Table~2 and the fit is given by the red line in
Figure~\ref{fig:fig12}.  Due to the large incompleteness in our last
magnitude bin, we have performed these fits to a limiting magnitude of
$g<21.65$ rather than $g<21.85$.  The errors on the parameters are
$\sigma_{\alpha}=0.2$, $\sigma_{\beta}=0.03$, $\sigma_{M^*}=0.09$,
$\sigma_{k1}=0.02$, $\sigma_{k2}=0.01$.

\begin{figure*}
\includegraphics[width=84mm]{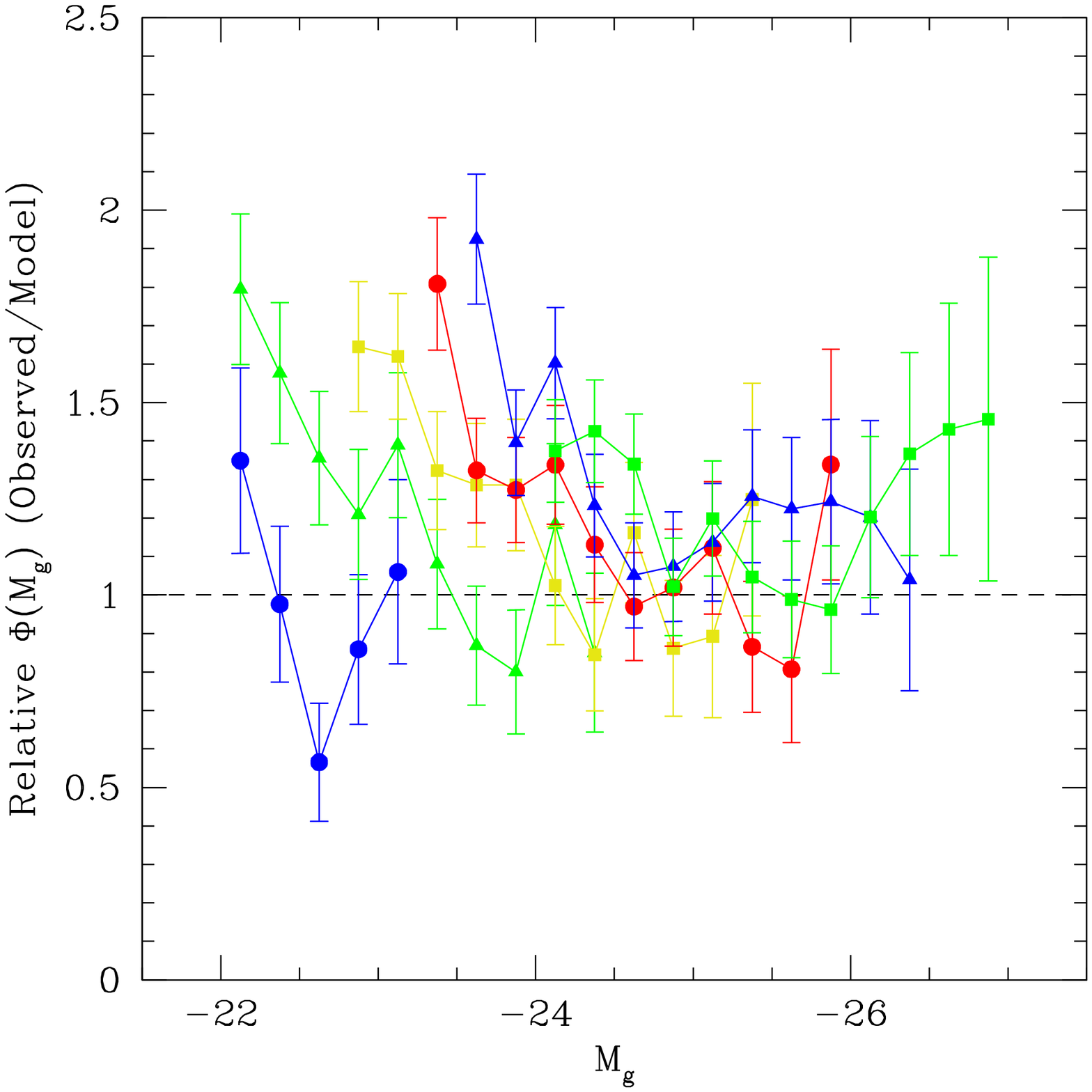}\hfil\includegraphics[width=84mm]{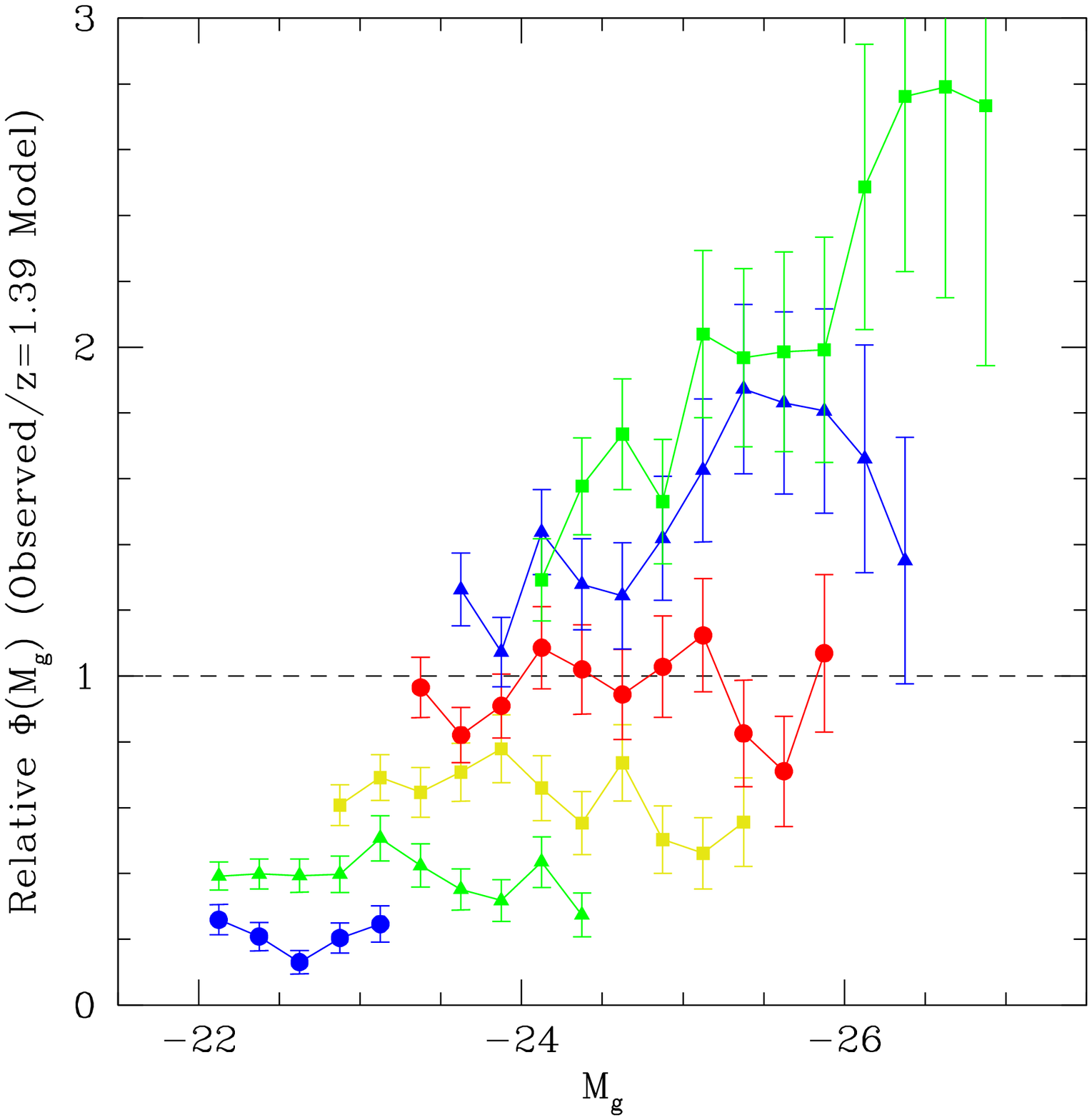}
\caption{{\em Left:} Ratio of luminosity function from
Fig.~\ref{fig:fig12} to the polynomial evolution models from
Croom et al.\ (2004)
. Colours and points are as in Figure~20 of
Croom et al.\ (2004) , specifically blue circles, green triangles,
yellow squares, red circles, blue triangles, and green squares,
corresponding to redshifts 0.54, 0.825, 1.11, 1.39, 1.67, and 1.955,
respectively.  Note the deviation from unity at the faint end in each
redshift bin.  {\em Right:} The ratio of the luminosity function to
our $z=1.39$ maximum likelihood model.}
\label{fig:fig13}
\end{figure*}

Since there are relatively few bright quasars in our sample to tie
down the bright end slope, we have also attempted to fix all of the
parameters except for the faint end slope ($\beta$) and the
normalization to those found by \nocite{csb+04}{Croom} {et~al.} (2004), specifically
$\alpha=-3.31$, $M_g^*=-21.61+5\log h_{70}$, $k_1=1.39$, and
$k_2=-0.29$.  The resulting faint end slope is then
$\beta=-1.45\pm0.03$ (with $\Phi^*=1.83\times10^{-6}
h_{70}^3$\,Mpc$^{-3}$\,mag$^{-1}$).  For both of these fits, a
$\chi^2$ comparison of this model to the $1/V_a$ data is formally
rejected; see Table~2.  We also note that there is apparently
substantial covariance between the parameters.  For example, there is
a significant difference in the faint end slopes of the \nocite{bsc+00}{Boyle} {et~al.} (2000)
and \nocite{csb+04}{Croom} {et~al.} (2004) analysis (as shown by the cyan and yellow lines in
Fig.~\ref{fig:fig12}), yet there is only a 1 per cent difference in
the total expected counts to the limiting magnitude of the 2QZ survey
($b_J=20.85$).  To the fainter limit of our survey, we find that the
final 2QZ parameterization \nocite{csb+04}({Croom} {et~al.} 2004) significantly underpredicts
(by 32 per cent) the total number of quasars to $g<21.65$, while the
\nocite{bsc+00}{Boyle} {et~al.} (2000) parameters yield a much better fit to the 2SLAQ data
(see Fig.~\ref{fig:fig12} and Table~\ref{tab:cum}).  The deviation
from the best fit 2QZ model can be seen better in the left-hand panel
of Figure~\ref{fig:fig13} where we have normalized our derived values
by the best fit polynomial evolution model from \nocite{csb+04}{Croom} {et~al.} (2004).  The
right hand panel is similar except that the data have all been
normalized to our $z=1.39$ model in order to better show the redshift
evolution of the quasars.  All of the above suggests that the adopted
parameterization is not the optimal one; however, it still has
considerable utility in terms of predicting counts of faint quasars
and as an input for theoretical models.

\begin{table*}
\begin{center}
\caption{Summary of maximum likelihood fits for the parameterization
adopted for the 2QZ analysis (double power-law with luminosity
evolution parameterized as a 2nd order polynomial in redshift) and our
adopted cosmology (\S~1).  The redshift limits is $0.4<z<2.1$, and
objects must be brighter then $M<-22.5$.  $N_Q$ indicates the number
of quasars per square degree expected for
$18.0<g<21.65$.}\label{tab:qlf}
\begin{tabular}{lcccccccccc}
\hline
Sample & $\alpha$ & $\beta$ & $M^*$ & $k_1$ & $k_2$ & $\Phi^*$ & $N_Q$ & $\chi^2$ & $\nu$ & $P_{\chi^2}$ \\
\hline
Boyle et al. (2000) & $-3.41$ & $-1.58$ & $-21.92$ & 1.36 & $-0.27$ & 9.88e-7 & 66.8 & & & \\
Croom et al. (2004) & $-3.31$ & $-1.09$ & $-21.61$ & 1.39 & $-0.29$ & 1.67e-6 & 54.4 & & & \\
2SLAQ + Croom et al. (2004) & $-3.31$ & $-1.45$ & $-21.61$ & 1.39 & $-0.29$ & 1.83e-6 & 83.8 & 161.5 & 55 & 2.1e-12 \\
2SLAQ only & $-3.28$ & $-1.78$ & $-22.68$ & 1.37 & $-0.32$ & 5.96e-7 & 79.8 & 149.0 & 51 & 1.5e-11\\
\hline
\end{tabular}
\end{center}
\end{table*}

We have also attempted to use the parameterizations of the luminosity
function that were used by \nocite{wwb+03}{Wolf} {et~al.} (2003) since
our data, like that of COMBO-17, appears to show less of a break in
the luminosity function than previous work.  The best fit forms and
parameters from \nocite{wwb+03}{Wolf} {et~al.} (2003) match the 2SLAQ
data over a limited range in redshift and absolute magnitude, but
these fits do not agree with the 2SLAQ data at the bright end and for
lower redshifts.  We were also unable to derive better fits to the
2SLAQ data using such parameterizations, likely because of the lack of
dynamic range at the bright end of the distribution.  However, it is
clear that other parameterizations, like those adopted by COMBO-17,
are worth pursuing.

\subsection{X-ray comparison}

\begin{figure}
\includegraphics[width=84mm]{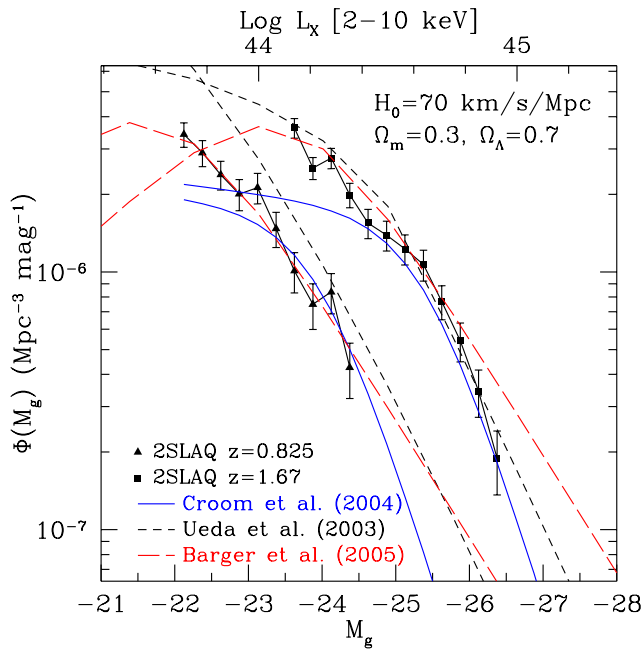}
\caption{Comparison of the 2SLAQ optical quasar luminosity function
with X-ray quasar luminosity functions from the literature.  Shown are
the $z=0.825$ ({\em triangles}) and $z=1.67$ ({\em squares}) QLF from
2SLAQ and the best fit models for those redshifts from three other
papers.  The models are from 
Croom et al.\ (2004)
({\em blue solid line}),
Ueda et al.\ (2003)
({\em dashed black line}) and 
Barger et al.\ (2005)
({\em red
long dashed line}).}
\label{fig:fig14}
\end{figure}

We can test the robustness of the faint end of the 2SLAQ luminosity
function by comparing to quasar luminosity functions derived from
X-ray selected samples which are thought to suffer less incompleteness
as a result of the dust-penetrating nature of X-ray photons.  In
Figure~\ref{fig:fig14} we compare the $z=0.825$ and $z=1.67$ redshift
bins from Figure~\ref{fig:fig12} to \nocite{csb+04}{Croom} {et~al.} (2004) and two quasar
luminosity functions derived from hard X-ray surveys
\nocite{uao+03,bcm+05}({Ueda} {et~al.} 2003; {Barger et al.} 2005).

In these comparisons, we have converted between $M_g$ and
$\log(L[2-10{\rm keV}])$ as follows.  First we take our K-corrected
$M_g$ and convert it to (rest-frame) $\log(L_g [{\rm erg\,s^{-1}\,cm^{-2}\,Hz^{-1}}])$ as prescribed by the
definition of an AB \nocite{og83}({Oke} \& {Gunn} 1983) magnitude and an absolute magnitude
(with an assumed distance of 10 pc)
\begin{equation}
\log(L_g) = -0.4(M_g - 5 + 48.6) + \log(4\pi) + 2\log(3.086\times10^{18}).
\end{equation}
Next we assume a power-law spectral index of $\alpha_{\nu}=-0.5$ to
convert from rest-frame $g$ (4669\AA) to rest-frame 2500\AA\
according to
\begin{equation}
\log(L_{2500}) = \log(L_g) - 0.5\log(4669/2500).
\end{equation}
We then extrapolate to $\log(L_{2keV})$ assuming a luminosity
dependent 2500\AA\ to 2\,keV slope, $\alpha_{ox}$, \nocite{vbs03}({Vignali}, {Brandt} \& {Schneider} 2003):
\begin{equation}
\alpha_{ox} = -0.11\log(L_{2500}) + 1.85
\end{equation}
and 
\begin{equation}
\log(L_{2keV}) = \log(L_{2500}) +
\alpha_{ox}\log\left(\frac{\nu_{2keV}}{\nu_{2500}}\right).  
\end{equation}
(Using the revised $\alpha_{ox}-L_{2500}$ relationship from
\nocite{sbs+05}{Strateva} {et~al.} 2005 yields somewhat fainter
X-ray luminosities [$\sim0.15$ dex] at the bright end of our sample and would slightly flatten the X-ray QLFs in Fig.~\ref{fig:fig14}.)
Finally, we compute a 2--10\,keV luminosity by integrating over the
2--10\,keV range assuming a photon index of $\Gamma=1.9$
($\alpha_x=-0.9$).  For comparison with \nocite{uao+03}{Ueda} {et~al.}
(2003) we further correct for the fraction of X-ray type II AGN
according to their Figure~8 and an optical type II fraction of 0.5,
which is roughly consistent with their Figure~9.  In our comparison
with the broad-line AGN luminosity function of \nocite{bcm+05}{Barger
et al.} (2005), we have treated their parameterization as if it were
for a 2--10 keV luminosity rather than a 2--8 keV luminosity (since we
are primarily concerned with the comparisons of the QLF slope), and we
have applied a correction factor of $0.5$ in the overall
normalization.  Furthermore, our comparison with
\nocite{bcm+05}{Barger et al.} (2005) differs somewhat from their
comparison with \nocite{csb+04}{Croom} {et~al.} (2004) in that
\nocite{bcm+05}{Barger et al.} (2005) converted the optical and X-ray
luminosities to bolometric luminosities in a manner which assumes a
constant $\alpha_{ox}$, whereas we assumed the luminosity-dependent
$\alpha_{ox}$ given above.  For both comparisons with X-ray QLFs we
have converted the parameterizations to the cosmology adopted herein.

For the sake of facilitating the comparison of optical QLFs to X-ray
QLFs, we note that, in the syntax used by \nocite{uao+03}{Ueda} {et~al.} (2003) in their
Equation~6 (and similar notation used by \nocite{bcm+05}{Barger et al.} 2005, Equation~1)
$\phi_M=A/2.5$, $\alpha=-(\gamma_1+1)$, and $\beta=-(\gamma_2+1)$,
where $\phi_M$, $\alpha$, and $\beta$ are defined as in \nocite{pet97}{Peterson} (1997),
Equation~11.33 (and similarly by \nocite{csb+04}{Croom} {et~al.} 2004, Equation~11).

In each case, the X-ray luminosity functions show less curvature in
the faintest 2SLAQ bins than does the best fit model from 2QZ.  This
comparison is not meant to be strictly quantitative since X-ray
selected samples are more sensitive to obscured quasars and the
conversion between $M_B$ and $L_X$ involves a number of tenuous
assumptions.  However, these comparisons confirm that the steeper
2SLAQ faint end slope, while based on large correction factors, is
quite reasonable.  In particular, the agreement with the results of
\nocite{bcm+05}{Barger et al.} (2005) is remarkable.

\section{Discussion and Conclusions}

We have compiled a sample of 5645 quasars with $18.0<g<21.85$ and
$z<3$ using imaging data from the SDSS and the spectra from the 2dF
facility at the AAT.  We find a clear turnover in the optical number
counts; a single power-law is not a good fit over the magnitude range
sampled.  For $20<g<20.6$, the 2SLAQ number counts show a slight (but
statistically insignificant) excess over previous surveys, but the
cumulative number counts are roughly consistent with the faintest
surveys to 22nd magnitude.

In terms of the luminosity function, we find good agreement with the
2QZ results from \nocite{csb+04}{Croom} {et~al.} (2004) at the bright end, but the faint end
2SLAQ data require a steeper slope (higher density of quasars) than
the 2QZ results from \nocite{csb+04}{Croom} {et~al.} (2004).  The previous 2QZ results from
\nocite{bsc+00}{Boyle} {et~al.} (2000) agree significantly better with 2SLAQ at the faint end.
The lack of a well defined characteristic luminosity and covariance
between the maximum likelihood parameters can explain the good
bright-end agreement between the parameterizations studied and the
faint-end disagreement between 2SLAQ and the final 2QZ results
\nocite{csb+04}{Croom} {et~al.} (2004).  Comparing to type I quasar luminosity functions
derived from X-ray samples suggests that the slope of the faint end of
the 2SLAQ QLF is more accurate than the extrapolated faint end slope of
\nocite{csb+04}{Croom} {et~al.} (2004).

An understanding of the quasar luminosity function is an important
ingredient for many different types of extragalactic investigations.
In particular, as has been stressed by those working with X-ray
selected samples, investigations that depend on the optical QLF
explicitly may need to be reconsidered as a result of recent revisions
in the luminosity function of unobscured AGNs (not to mention obscured
AGNs).  Many investigations have an explicit dependence on the optical
QLF, for example \nocite{bsk+01}{Bianchi}, {Cristiani} \& {Kim} (2001) in their analysis of the UV
background; \nocite{hct02}{Hamilton}, {Casertano} \&  {Turnshek} (2002) in their estimate of the quasar host galaxy
luminosity function; \nocite{yt02}{Yu} \& {Tremaine} (2002) in their investigation of the growth
of black holes; \nocite{cbl+02}{Croom} {et~al.} (2002) and \nocite{wmd+04}{Wake} {et~al.} (2004) regarding the
clustering of AGN; \nocite{ogu03}{Oguri} (2003) in his determination of the expected
number of lensed quasars; \nocite{rsp+04}{Richards} {et~al.} (2004b) in their assessment of the
lensing probability of the most luminous high-redshift quasars; and
\nocite{fnl+01}{Fan} {et~al.} (2001a) in terms of the evolution of quasars from $z=0$ to
$z=6$.  The QLF has taken on even greater importance in recent years
with the realization that most massive galaxies host supermassive
black holes, the correlation between black hole mass and stellar
velocity dispersion \nocite{mtr+98,fm00,gbb+00}(e.g. {Magorrian} {et~al.} 1998; {Ferrarese} \& {Merritt} 2000; {Gebhardt} {et~al.} 2000), and the
possibility that feedback from quasars may play a role in the
evolution of galaxies in general \nocite{beg04}(e.g. {Begelman} 2004).  In
particular, models like those of \nocite{kh00}{Kauffmann} \& {Haehnelt} (2000), \nocite{wl02}{Wyithe} \& {Loeb} (2002), and
others that attempt to explain the evolution of galaxies and quasars,
rely on comparison with accurate observational determinations of the
QLF.

In fact, although the optical QLF presented herein is arguably the
most robust determination to date for a large optically selected
sample, for many applications an X-ray or far-IR QLF is more
appropriate.  That said, the luminosity function of optically selected
quasars will remain an important tool for extragalactic astronomy.
The primary reason for this is the sheer size of the optical quasar
sample (likely over 300,000 in the current SDSS imaging data alone;
\nocite{rng+04}{Richards} {et~al.} 2004a).  While the deepest X-ray surveys may uncover
thousands of AGNs per square degree, they do so over only a fraction
of a square degree and the sum total area of the sky covered by both
{\em Chandra} and {\em XMM-Newton} is unlikely to ever exceed even 1
per cent.  IR surveys with {\em Spitzer} will cover a somewhat larger
area than X-ray surveys, but not at nearly the same space density as
in the X-ray or with nearly the same total number as in the shallower,
but much wider optical surveys.  Thus, this sample of faint quasars
and the luminosity function derived from it will continue to provide
important inputs to future extragalactic investigations such as the
Dark Energy Survey (DES) and the Large Synoptic Survey Telescope
(LSST).

\section*{Acknowledgements}

Funding for the creation and distribution of the SDSS Archive has been
provided by the Alfred P. Sloan Foundation, the Participating
Institutions, the National Aeronautics and Space Administration, the
National Science Foundation, the U.S. Department of Energy, the
Japanese Monbukagakusho and the Max Planck Society. The SDSS Web site
is http://www.sdss.org/.  The SDSS is managed by the Astrophysical
Research Consortium (ARC) for the Participating Institutions. The
Participating Institutions are The University of Chicago, Fermilab,
the Institute for Advanced Study, the Japan Participation Group, The
Johns Hopkins University, the Korean Scientist Group, Los Alamos
National Laboratory, the Max-Planck-Institute for Astronomy (MPIA),
the Max-Planck-Institute for Astrophysics (MPA), New Mexico State
University, University of Pittsburgh, University of Portsmouth,
Princeton University, the United States Naval Observatory and the
University of Washington.  Spectroscopic observations were performed
with the 2dF instrument on the Anglo-Australian Telescope and we thank
the staff of the Anglo-Australian Observatory for their support.  We
thank Michael Weinstein and Michael J.\ I.\ Brown for assistance with
code development.  D.~P.~S. and D.~E.~VB were supported in part by NSF
grant AST-0307582.  M.~A.~S. was supported in part by NSF grant
AST-0307409.




\clearpage

\label{lastpage}

\end{document}